\documentclass[manuscript]{acmart}

\AtBeginDocument{%
  }

\setcopyright{acmlicensed}
\copyrightyear{2018}
\acmYear{2025}
\acmDOI{XXXXXXX.XXXXXXX}

\acmJournal{JACM}
\acmVolume{37}
\acmNumber{4}
\acmArticle{111}
\acmMonth{12}

\usepackage{subcaption}
\usepackage{pifont}

\begin{document}

\title{Tracing the Data Trail: A Survey of Data Provenance, Transparency and Traceability in LLMs}

\author{Richard Hohensinner}
\email{richard.hohensinner@pro2future.at}
\orcid{0009-0003-2820-5972}
\authornotemark[1]
\affiliation{%
  \institution{Pro2Future GmbH}
  \city{Graz}
  \state{Styria}
  \country{Austria}
}
\author{Belgin Mutlu}
\email{belgin.mutlu@pro2future.at}
\orcid{}
\authornotemark[1]
\affiliation{%
  \institution{Pro2Future GmbH}
  \city{Graz}
  \state{Styria}
  \country{Austria}
}

\author{Inti Gabriel Mendoza Estrada}
\email{inti.mendoza@openmaind.ai}
\orcid{}
\authornotemark[2]
\affiliation{%
  \institution{openmaind FlexCo}
  \city{Graz}
  \state{Styria}
  \country{Austria}
}

\author{Matej Vukovic}
\email{matej.vukovic@pro2future.at}
\orcid{}
\authornotemark[1]
\affiliation{%
  \institution{Pro2Future GmbH}
  \city{Graz}
  \state{Styria}
  \country{Austria}
}

\author{Simone Kopeinik}
\email{skopeinik@know-center.at}
\orcid{}
\authornotemark[3]
\affiliation{%
  \institution{Know Center Research GmbH}
  \city{Graz}
  \state{Styria}
  \country{Austria}
}

\author{Roman Kern}
\email{rkern@tugraz.at}
\orcid{}
\authornotemark[4]
\affiliation{%
  \institution{Graz University of Technology}
  \city{Graz}
  \state{Styria}
  \country{Austria}
}

\renewcommand{\shortauthors}{Hohensinner et al.}

\begin{abstract}

Large language models (LLMs) are deployed at scale, yet their training data life cycle remains opaque. 
This survey synthesizes research from the past ten years on three tightly coupled axes: (1) data provenance, (2) transparency, and (3) traceability, and three supporting pillars: (4) bias \& uncertainty, (5) data privacy, and (6) tools and techniques that operationalize them.
A central contribution is a proposed taxonomy defining the field's domains and listing corresponding artifacts.
Through analysis of 95 publications, this work identifies key methodologies concerning data generation, watermarking, bias measurement, data curation, data privacy, and the inherent trade-off between transparency and opacity.



\end{abstract}


\maketitle

\section{Introduction}
\label{chap_intro}

\begin{figure}[h]
  \centering
  \includegraphics[width=0.75\linewidth]{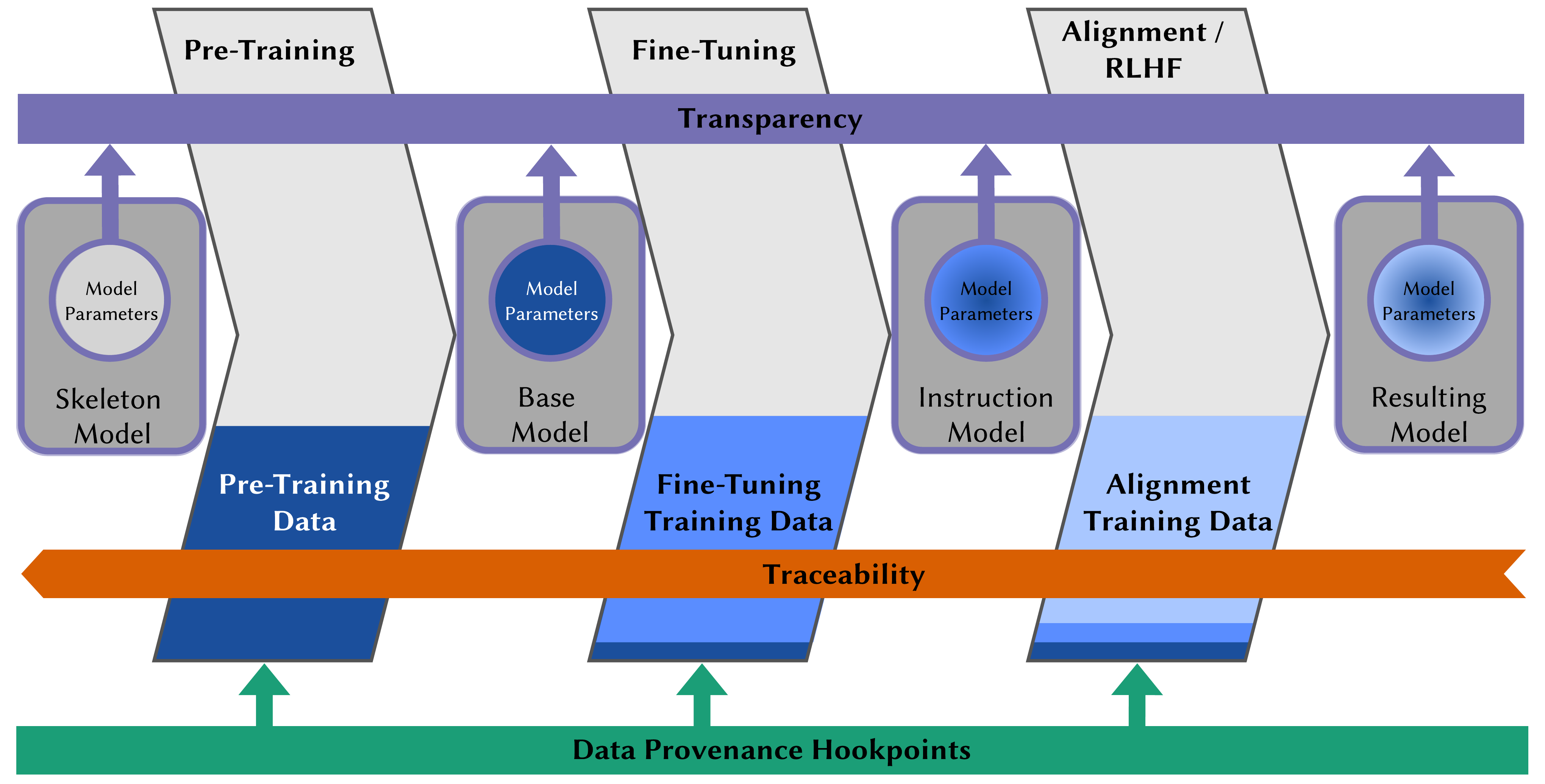}
  \caption{Illustration of data provenance (Chapter~\ref{chap_dp}), traceability (Chapter~\ref{chap_trace}), and transparency (Chapter~\ref{chap_trans}) linked to the data flow in LLMs. The supporting pillars i.e., bias, uncertainty (Chapter~\ref{chap_bias}), and privacy (Chapter~\ref{chap_privacy}) are directly linked to the training data sources, and provenance tools and techniques (Chapter~\ref{chap_tools}) enhance the data cycle at various intersections.}
  \Description{Illustration of data provenance (Chapter~\ref{chap_dp}), traceability (Chapter~\ref{chap_trace}), and transparency (Chapter~\ref{chap_trans}) linked to the data flow in LLMs. The supporting pillars i.e., bias, uncertainty (Chapter~\ref{chap_bias}), and privacy (Chapter~\ref{chap_privacy}) are directly linked to the training data sources, and provenance tools and techniques (Chapter~\ref{chap_tools}) enhance the data cycle at various intersections.}
  \label{fig_one}
\end{figure}

Modern Large Language Models (LLMs) are built with large amounts of data. 
This data stems from various sources, including crawled textual content of the Web, like the FineWeb dataset \cite{10.5555/3737916.3738886}, containing up to 15 trillion tokens (44 Terabytes) of refined text. 
However, LLMs are currently incapable to trace back the generated output to the original data sources.
Even worse, major LLM creators keep the data origins and training processes of their flagship GPT models secret and create opaque black-box models as a result, motivated by economic considerations and proprietary concerns.
These principles force AI researchers to operate under uncertainty regarding data origins and their ensuing implications.

In this work, we investigate the concepts of data provenance, transparency, and traceability linked to the domain of LLMs. Furthermore, we examine the linkages of three supporting pillars to the data origins of LLMs: bias, privacy, and provenance tools \& techniques. 
Figure~\ref{fig_one} provides an overview of these domains based on the life cycle of an LLM.
%
%
%
Current rapid developments in LLMs lead to huge increases in model parameter size. 
Looking at developments of recent years, OpenAI's GPT-1 \cite{radford2018improving} started with 117 million parameters in 2018, followed by GPT-2 \cite{radford2019language} with 1,5 billion parameters in 2019, and GPT-3 \cite{brown2020language} in 2020 with 175 billion parameters. 
In 2025, flagship models like DeepSeek-R1 \cite{guo2025deepseek} are already at 671 billion parameters, and Moonshot AI's Kimi K2 \cite{kimiteam2025kimik2openagentic} further exceeds this with 1 trillion parameters in total.
This skyrocketing growth of model parameters not only demands an increased amount of computing power for deployment but also requires a corresponding growth of training data to be exploited effectively. 
Furthermore, with the rising need for data, data quality has to be maintained accordingly to avoid the use of corrupted data to decrease the risk of hallucinations caused by low-quality data sources \cite{ma-etal-2025-understanding, gautam975impact}.
This raises questions on how this hunger for data can be satisfied sustainably and how the provenance of training data can aid in mitigating LLM-related issues, such as hallucinations. 
To effectively utilize this provenance knowledge, LLMs need to possess a certain degree of transparency.
This transparency allows investigations and sense-making of internal states and decision-making of these systems.
Additionally, transparency measures for LLMs enable tracing data and knowledge flows through LLMs.
Traceability, on the other hand, provides the ability to link underlying source data to generated outputs.
However, tracing information through LLMs is complicated by the intrinsic decoupling of language as words, which exist only as complex mathematical representations within these systems.


The aforementioned increasing demand for high-quality data far exceeds the scope of manually human-verifiable sources, and new ways of quality assessment have to be found.
Filtering huge amounts of datasets from the web, but keeping a high threshold of quality, is a difficult task, tedious to do manually, yet treacherous to automate securely. 
A strand of research suggests using LLMs and their data generation capabilities themselves to automate data annotation, curation, and filtering \cite{Choi2024, Rashid2020, Golany2024}. 
However, further processing data that originates from a system whose own data sources are not fully disclosed may introduce a new range of potentially harmful problem sources.

\begin{figure}[h]
  \centering
  \includegraphics[width=0.8\linewidth]{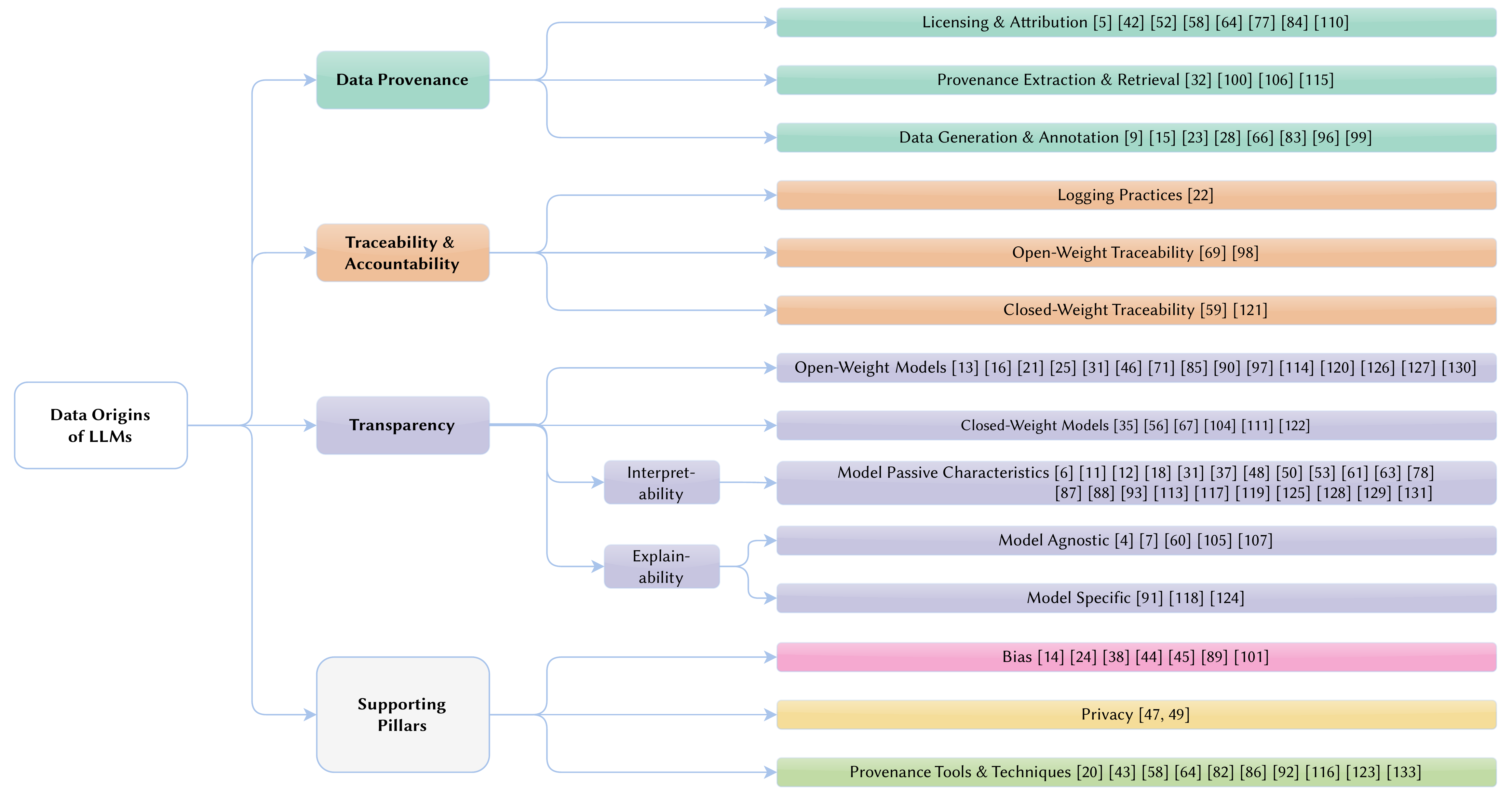}
  \caption{Taxonomy of our systematic literature review, outlining the three main axes - Data Provenance, Transparency, and Traceability - together with their supporting pillars. Bias \& Uncertainty, Data Privacy, and Provenance Tools \& Techniques provide additional context for characterizing data origins of LLMs.}
  \Description{Taxonomy of our systematic literature review, outlining the three main axes - Data Provenance, Transparency, and Traceability - together with their supporting pillars. Bias \& Uncertainty, Data Privacy, and Provenance Tools \& Techniques provide additional context for characterizing data origins of LLMs.}
  \label{fig:taxonomy}
\end{figure}

Data provenance regarding LLMs involves tracking the origins of data sources.
This includes all data that is ingested into LLMs, i.e., through training, fine-tuning, or alignment processes, as well as settings and restriction sources.
Tracing data through a language model is considered difficult because the data involved is no longer directly accessible within the model after training. 
Language models do not store data in their conventional state - they store the respective knowledge and links from data in their parameters, also referenced as weights, in a mathematical representation, which impedes tracing information conventionally~\cite{morris2025languagemodelsmemorize}.

The overarching goal of traceability research in this domain is to establish an end-to-end traceable data flow pipeline, from the user query and the system's response to the knowledge origins in the initial training data.
Recent work aims to analyze the biology of LLMs by investigating their inner circuits in an interpretive study \cite{lindsey2025biology} to further understand the model's decision-making process. 
A deeper understanding of these circuits could lead to more profound insights into the connections between the initial training data and the inner mechanisms of a language model. 
Thus, investigations in our work aim to tackle the traceability of training data and their origins for LLMs. 

Further, the issue of using LLM-generated content as a training basis for new models is a timely, pressing topic. 
More and more models are currently being trained on data generated by generative AI.
We want to emphasize that this procedure contains many risks, such as biases propagating across language models, proprietary data that was ingested unintentionally by a donor (or teacher) model being forwarded to a receiver model, and the uncertainty of unknown data sources growing to a second layer.

Throughout the training process, the intrinsic knowledge of the training data is transferred into the model's parameters. 
However, the model parameters do not link back to the training data by default. 
This creates a transparency gap, if only the resulting language model, but not its training sources, is accessible to users and experts. 
Furthermore, opening questions of where particular knowledge from the source data is stored within the model parameters, and whether it is possible to detect particular parameter regions that can be traced back to originating data sources. 
These questions can only be investigated under the conditions that model parameters and their respective sources are fully disclosed.

LLMs with transparent architecture and accessible weights are referred to as \textit{open-weight} models. 
In contrast, language models that do not provide access to their internal states are differentiated as \textit{closed-weight} models. 
With this gap spreading from fully open to entirely opaque, naturally, there are LLMs with transparency measures in between, disclosing only parts of their internal states.
Additionally, the grade of access also decides the ability to run these language models on local infrastructure. 
The ability to host a language model on-premises opens opportunities for in-depth research on model parameters in regard to interpretability and explainability. 
With interpretability being a passive property of a model that defines to what extent the internals of LLMs can be \textit{directly understood} by humans, and explainability determining \textit{active processes} which produce human-understandable reasons and decision-making of LLMs.
%
%
%
We found that data provenance, traceability, and transparency go hand in hand in the domain of LLMs, as it is impossible to trace data flows through non-transparent, opaque systems, and similarly impossible to match data origins to final results without the means of traceability.

Furthermore, LLMs are prone to taking on bias from data sources \cite{Taori2022}, because the sheer amount of data involved in the training of current language models, as well as the lack of knowledge and context, make it difficult to mitigate.
In systems that interact with humans, bias can never be fully eliminated, only mitigated to a certain degree.
This mitigation usually relates to the application context, and the general-purpose nature of LLMs renders this task particularly hard to counteract.
Research on data provenance for LLMs could help reduce unwanted bias in models by attributing detected bias to specific parts of the training data. 
Succeeding training runs could then be trained with bias-annotated datasets in future iterations to improve over time. 
Provenance information also enables differentiating between bias inherited from data sources and the intrinsic bias of the language model architecture or training procedure. 
Additionally, provenance-driven data investigations for LLMs can help to tackle cases of misinformation or other harmful content, by following the data trail back to the flawed data sources. 

LLMs are currently considered to be in a gray zone of privacy legislation. 
However, LLM creators are increasingly under pressure due to data privacy and protection laws such as the European AI-Act or the General Data Protection Regulation (GDPR). 
Alongside this, one particular data protection right, the "Right to be forgotten", becomes increasingly important for the domain of LLMs, especially due to their final states and lack of post-hoc editing. 
Although research already investigates methods to edit individual model parameters, e.g., deleting certain information \cite{hsueh2024editingmindgiantsindepth}, there is no general solution yet to erase one's erroneously digested data from a trained language model. 
We found that current research, although of utmost importance, seldom mentions the topic of data privacy concerning LLM data usage.

Finally, current research already provides, to some extent, aid to tackle the opaque data flow into LLMs.
We investigate works for their applicability to the domains of data provenance, traceability, and transparency, and report provenance tools and techniques that may help the cause and inspire future research.


\subsection{Related Surveys}
\label{chap_relsuvs}

One can observe an increased interest in topics related to data provenance and related concepts, highlighted by a surge of articles and surveys in the past year.
A chronologically sorted overview of related surveys is given in Table~\ref{tab:relsurvs}.
Older works \cite{simmhan2005surveytechniques, herschel2017survey} focused on data collection and generation, while newer works \cite{hu2020survey, simmhan2005survey, ahmed2023data, pan2023data} are shifting towards the relationship between the data and data processing.
The most recent trend here is a focus on LLMs \cite{gregori2025llm, zhou2025survey, pang2025large}.

Simmhan et al. \cite{simmhan2005surveytechniques} present an overview of techniques for data provenance, stating that provenance (in 2005) is still in an exploratory field with several open questions.
Herschel et al. \cite{herschel2017survey} investigate where provenance information stems from, in what states it exists, and what this information can be used for.
Additional surveys explore data provenance in domains such as the Internet of Things (IoT) \cite{hu2020survey}, e-Science \cite{simmhan2005survey}, healthcare \cite{ahmed2023data}, and security \& privacy \cite{pan2023data}.
In 2025, Gregory et al.\cite{gregori2025llm} investigate how LLMs can be used for the collection and management of data provenance information, Zhou et al. \cite{zhou2025survey} analyze the interconnections of LLMs and the overarching domain of data management, and Pang et al. \cite{pang2025large} present an overview of the different perspectives from which LLMs source their information.

Our survey follows this trajectory and increases the scope by investigating the capabilities of different types of LLMs, the means of transparency with a separate view on interpretability \& explainability, and in addition covering provenance tools \& techniques for LLMs.

\begin{table*}
  \caption{Overview of related surveys, highlighting the shift in interest from data collection to data processing and finally LLMs.}
  \label{tab:relsurvs}
  \scalebox{0.70}{
  \begin{tabular}{lclcccccccccc}
    \toprule
    \textbf{Survey} & \textbf{Year} & \textbf{Focus} & \textbf{LLM} & \textbf{Data Prov.} & \textbf{Trace.} & \textbf{Trans.} & \textbf{Interp.} & \textbf{Expl.} & \textbf{Bias} & \textbf{Priv.} & \textbf{Pr.~Tools} & \textbf{Pr.~Techn.}\\ 
    \midrule
    \cite{simmhan2005surveytechniques} & 2005 & Data Prov. Techniques               & $\times$   & \checkmark & $\times$   & $\times$   & $\times$   & $\times$   & $\times$   & $\times$   & $\times$   & $\times$  \\
    \cite{herschel2017survey}          & 2017 & Data Provenance             & $\times$   & \checkmark & \checkmark & $\times$   & $\times$   & $\times$   & $\times$   & \checkmark & $\times$   & $\times$  \\
    \cite{hu2020survey}                & 2019 & Data Prov. in IoT                   & $\times$   & \checkmark & \checkmark & $\times$   & $\times$   & $\times$   & $\times$   & \checkmark & $\times$   & $\times$  \\
    \cite{simmhan2005survey}           & 2020 & Data Prov. in e-Science             & $\times$   & \checkmark & $\times$   & $\times$   & $\times$   & $\times$   & $\times$   & $\times$   & $\times$   & \checkmark\\
    \cite{ahmed2023data}               & 2023 & Data Prov. in Healthcare            & $\times$   & \checkmark & \checkmark & $\times$   & $\times$   & $\times$   & $\times$   & \checkmark & $\times$   & $\times$  \\
    \cite{pan2023data}                 & 2023 & Data Prov. in Sec. \& Priv.         & $\times$   & \checkmark & \checkmark & $\times$   & $\times$   & $\times$   & $\times$   & \checkmark & $\times$   & $\times$  \\
    \cite{gregori2025llm}              & 2025 & Col. \& Mgmt. of Data Prov.         & \checkmark & \checkmark & \checkmark & \checkmark & \checkmark & \checkmark & \checkmark & $\times$   & $\times$   & $\times$  \\
    \cite{zhou2025survey}              & 2025 & LLM \& Data Mgmt.           & \checkmark & \checkmark & \checkmark & \checkmark & $\times$   & $\times$   & \checkmark & \checkmark & \checkmark & \checkmark\\
    \cite{pang2025large}               & 2025 & LLM Sourcing                & \checkmark & \checkmark & \checkmark & \checkmark & \checkmark & $\times$   & \checkmark & \checkmark & $\times$   & \checkmark\\
     Ours                       & 2025 & Data Prov., Trace. \& Trans.        & \ding{51}  & \ding{51}  & \ding{51}  & \ding{51}  & \ding{51}  & \ding{51}  & \ding{51}  & \ding{51}  & \ding{51}  & \ding{51}\\

    \bottomrule
  \end{tabular}}
\end{table*}

\subsection{Contribution}
\label{chap_contri}

\begin{figure}
  \centering
  \includegraphics[width=0.75\linewidth]{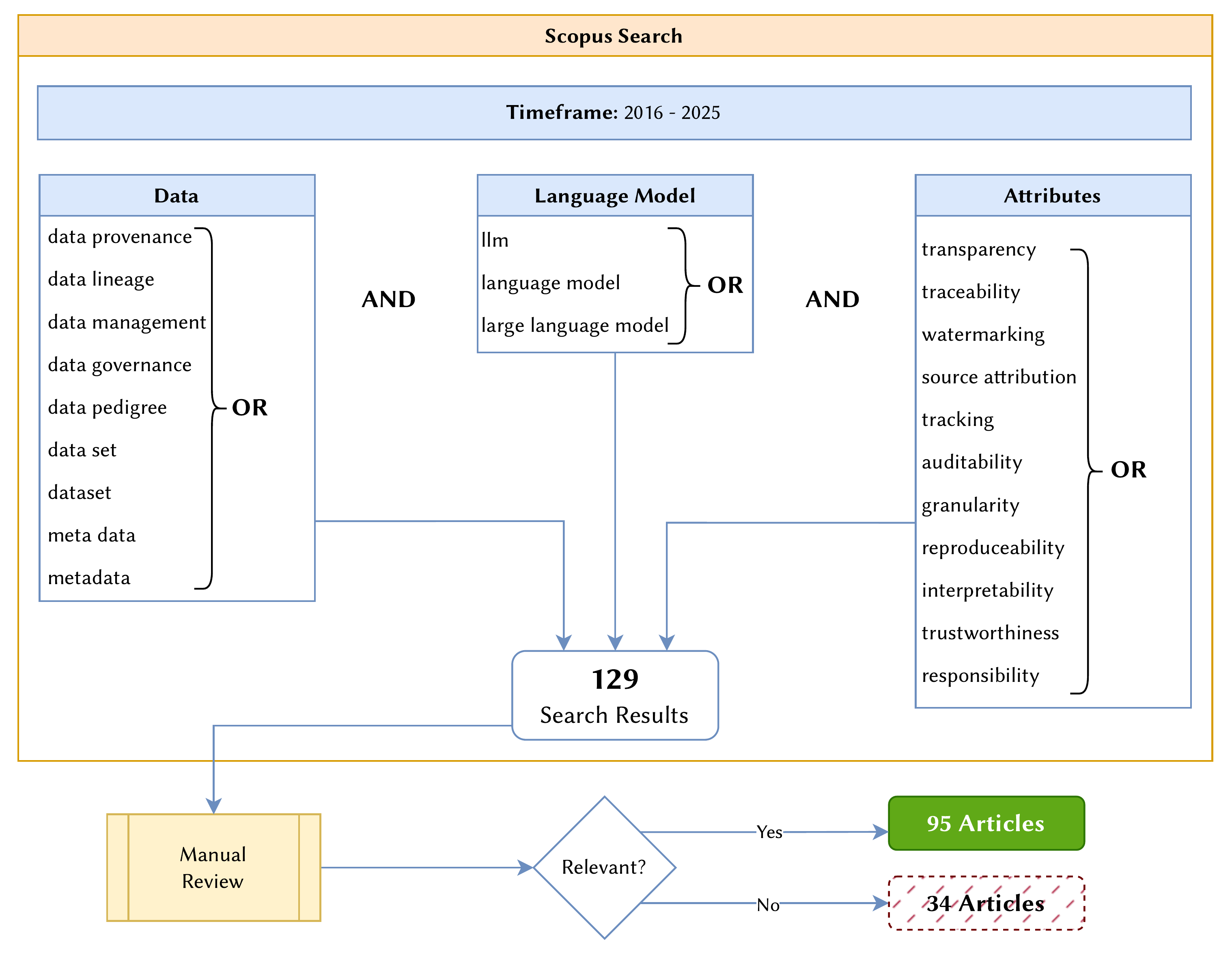}
  \caption{Illustration of the Scopus search, query construction, and manual reviewing process. The timeframe 2016 - 2025 spans over our three filters: Data, Language Model, and Attributes, leading to 129 search results.}
  \Description{Illustration of the Scopus search, query construction, and manual reviewing process. The timeframe 2016 - 2025 spans over our three filters: Data, Language Model, and Attributes, leading to 129 search results.}
  \label{fig:scopus_scan}
\end{figure}

This survey provides an analysis of the data origins of LLMs, setting a particular focus on data provenance, traceability, and transparency.
This includes traceability approaches such as watermarking and transparency investigations regarding the model type, as well as interpretability and explainability of LLMs. 
Additionally, supporting pillars such as bias, data privacy, and provenance tools \& techniques are analyzed.
Although relevant to the domain of data governance, topics related to data security and data quality are excluded from this survey as we deem they deserve their own standalone survey.
Our main contributions are related to a retrospective analysis, the current state of the art, and a future outlook:

\begin{enumerate}
\item[(1)] Providing a thorough literature review on LLM research, concentrating on data usage from the past 10 years, with each article containing at least one of the three criteria: 1) data usage for LLMs, 2) data generation for and by the usage of LLMs, 3) data provenance in the field of machine learning and AI.
\item[(2)] Creating an overview of the current landscape of approaches and their limitations, and establishing a new research domain by exploring the LLM's environments for data provenance capabilities.
\item[(3)] Identifying current research gaps and proposing future research directions for data provenance, traceability, and transparency for LLMs and data-driven explainable AI.
\item[(4)] Establishing guidelines as highlighted by selected work, and frameworks (Chapter~\ref{chap_tools}) in the context of this survey.
\end{enumerate}

\subsection{Survey Methodology}
\label{chap_method}


\begin{figure}[t]
    \centering
    \subcaptionbox{Distribution of articles to the main pillars: data provenance, transparency and traceability, and supporting pillars: bias \& uncertainty, privacy and provenance tools \& techniques.}{\includegraphics[width=0.45\textwidth]{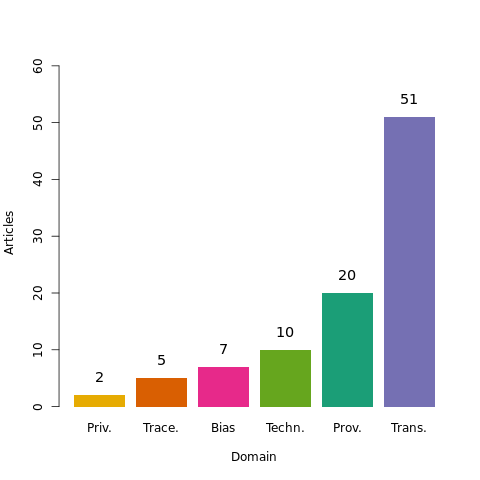}}%
    \hfill 
    \subcaptionbox{Distribution of articles and domains by year. The low count for 2025 is due to the March cutoff date; interest is expected to continue growing in line with recent trends.}{\includegraphics[width=0.45\textwidth]{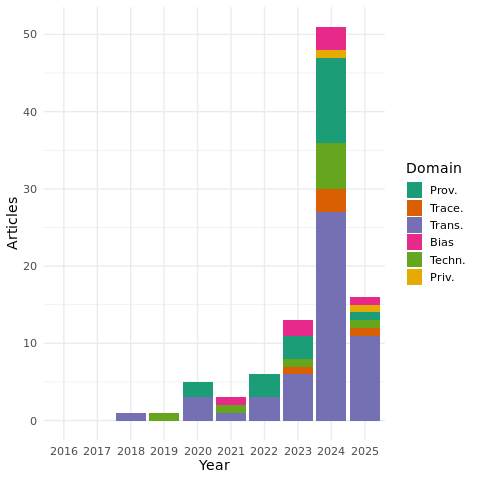}}%
  \caption{Overview of the considered articles and their domains.}
  \label{fig:worksperyear}
  \Description{Overview of the considered articles and their domains.}
\end{figure}

To achieve reproducibility in the considered articles, we opted for a systematic review based on a query and a scientific database for retrieval, where we utilized Scopus\footnote{https://www.scopus.com/}.
Figure~\ref{fig:scopus_scan} outlines the search process and query keywords.
The considered time span was set to cover the last ten years of research related to LLMs, spreading from January 2016 to March 2025 (date of search). 
Next, we applied three additional filters during the advanced Scopus search, all of which needed to be met to make it into the pool of articles. 
The first filter ensured a focus on data (provenance) and related data-centered topics, such as data lineage, governance, and management, which are terms frequently used in connection with data provenance.
The second filter ensured the usage of Large Language Models (LLMs), written in variations, so as not to miss any work using just one writing style. 
And, the third filter incorporated quality attributes like transparency, traceability, auditability, or reproducibility, which stemmed from our preliminary literature review on the topic of data origins of LLMs.
This search yielded a total of 129 results that matched our criteria.
We then manually processed these 129 search results.
The search results also included thirteen complete conference proceedings. 
We filtered these proceedings for duplicates and extracted ten additional documents for review.
Whereas the initial conference proceedings documents were excluded.
Finally, we further filtered the articles via several criteria, where at least one of three conditions needed to be met for relevancy: 
(1) Involvement of data usage in connection with LLMs, 
(2) data generation for and from the usage of LLMs, 
and (3) data provenance in the field of machine learning (ML) and artificial intelligence (AI).
We applied these criteria, resulting in the removal of 34 papers, which were considered irrelevant.
Thus, the final set of articles comprises 95 articles, which we further grouped into sub-domains.
Starting from the central topic of data provenance, we found that traceability and transparency are closely linked to the data origins of LLMs. Additional investigation revealed that bias, uncertainty, privacy, tools, and techniques are supporting pillars of the field. The distribution of articles to these domains is visualized in Figure~\ref{fig:worksperyear}.
Furthermore, our detailed grouping of the domains, illustrated in Figure~\ref{fig:taxonomy}, resulted in the following structure of this survey:
Chapter~\ref{chap_dp}: Data Provenance \& Attribution, Chapter~\ref{chap_trace}: Traceability \& Accountability, Chapter~\ref{chap_trans}: Transparency \& Opacity, Chapter~\ref{chap_bias}: Bias \& Uncertainty, Chapter~\ref{chap_privacy}: Data Privacy, and Chapter~\ref{chap_tools}: Provenance Tools \& Techniques.

\section{Preliminaries}

Data provenance for LLMs is closely linked to traceability and transparency. 
Means that support the traceability of information are key to maintaining data provenance throughout complex LLM pipelines and to mitigating obfuscation through parameterization. 
The concept of transparency, on the other hand, lays the foundation for LLM data provenance, as it enables the observation of LLM pipelines and their data flows. Furthermore, transparency incorporates the disciplines of explainable and interpretable AI. 
Also of utmost importance is the scientific discourse on biases that are inherent in data sources and are propagated into LLMs through those during training.
Provenance tools and techniques tackle the opaque landscape of data that flows through LLMs with the goal of establishing understandable model decision-making processes.
Whereas data privacy concerns protecting one's identity, personal information, and proprietary data sources.
Investigating data provenance for LLMs can help identify and address existing problems, providing a quintessence for more transparent and traceable data handling.

\subsection{Scope of LLMs and data sources}
We consider LLMs at all stages, including foundational (or basic) models, fine-tuned (or instruction) models, and models that underwent alignment with Reinforcement Learning with Human Feedback (RLHF). 
The training data of LLMs is hard to classify in general. 
The same dataset that was used to pre-train one model could also be used in another model's fine-tuning step.
Additionally, data sources may change over time through manual editing or complex data manipulation, and prior iterations may become inaccessible over time. 
Thus, we differentiate the \emph{data flow} into language models as training data (pre-training, fine-tuning, alignment) for parameterization and inference input (user queries). 
The former has a direct impact on the language model's weights, whereas inference input does not change an LLM's internal memory. 
Furthermore, we draw a distinct boundary between the inside and outside of the LLM system. 
Outside of the system: (Training) Data sources, user prompts \& queries, and outputs as tokens and logprobs (logarithmic probabilities). 
Inside the system: Model weights, also known as parameters, layers, and the architecture.
Additional grey zones exist, such as pre-processing methods of prompts before they are transferred into the LLM, or post-hoc filters applied to generated outputs, and post-processing of responses for the User Interface. 
We consider these methods as grey zones, because they may or may not be implemented in LLM systems and their environment.

\subsection{Data Provenance}

\begin{quote}
``Data provenance, a record that describes the origins and processing of data, offers new promises in the increasingly important role of artificial intelligence (AI)-based systems in guiding human decision making.'' \cite{Werder2022}
\end{quote}

Data provenance is mainly focused on the origin and source of a data object, dating back to its original creation. 
Usually, this provenance information is documented and stored within additional metadata. 
In more complex environments, so-called provenance graphs \cite{WANG2025125877} are used to document the processing of data and information flow.
Data Lineage, often mistakenly used as a synonym for data provenance, is focused on the transformation of data over time. 
A Wikipedia entry can be used as an example, as it forms a piece of information that is continuously updated by multiple authors over time.
Also, a configuration file stored on a server, accessed by multiple system administrators, can be used as an example. 
Furthermore, datasets that change in iterations, either through extension or reduction, are also subject to data lineage.
Similar to the provenance information, data lineage is stored as metadata alongside the data objects, and changes can be visualized by diagrams over time. 
We formalize the key difference in both terms as follows:
\begin{itemize}
  \item \textbf{Data Provenance:} Keeps track of the origin of data sources and their creation.
  \item \textbf{Data Lineage:} Keeps track of the transformation of data sources over their entire life cycle.
\end{itemize}

Regarding LLMs, the concept of data provenance can be used to shine light into the obfuscated entanglement that is created by billions of parameters coexisting in modern language models. 
This could be achieved by tracing the origin of a generated token, via the model's weights, all the way back to its initial training data source evidence. 

For this, two approaches exist: (1) Keeping track of original provenance information throughout the parametrization and generation process and (2) Searching for the provenance sources post-hoc, using the model's output and the entirety of involved data sources \cite{liu2025}.

The concept of data provenance in the field of LLMs seems straightforward, yet, opens many questions. 
Throughout the rapid growth of LLMs in the field of AI in recent years, primary focus was put on metrics like performance, speed for faster inference, range of tasks through parameter size, adherence to social norms by avoidance of generating bias and harmful content, and evaluation metrics for domain benchmarks, e.g., coding or solving mathematical problems. 
Although quintessential for the widespread success of LLMs, the underlying dependency on high-quality data and its corresponding provenance appears to be neglected.

\subsubsection{Granularity and Linkages}

We section data source levels for this work as: (1) Open web (Big Data), (2) datasets, (3) documents \& files, (4) document chunks, (5) sentences, (6) tokens (\& embeddings).
And, implicit source linkages as: (1) Crawling \& curation (from web to datasets), (2) document chunking, i.e., sliding windows (documents to text chunks, and (3) embeddings (sentences to tokens).
In addition, we note that multimodal data, that is more complex than pure text, requires additional data processing steps. Images and video data require visual embeddings for Large Visual Language Models (LVLMs), and audio sources can be preprocessed with text-to-speech or speech-to-text, respectively.

\subsection{Traceability}

\begin{quote}
``Data traceability is the actual exercise to track access, values, and changes to the data as they flow through their lineage. Data traceability can be used for data validation and verification as well as data auditing. [...] Data traceability might also be required in an auditing project to demonstrate transparency, compliance, and adherence to regulations'' \cite{Allen2015}
\end{quote}

Traceability is concerned with understanding the connections between data artifacts and transformations throughout the LLM pipeline and understanding the inner workings of the LLM chain, such as attention heads, layers, and parameter circuits.
Through this, traceability forms a crucial foundation for data provenance - and, thus, is heavily linked to it.

New approaches, such as watermarking \cite{Ye2025}, fingerprinting \cite{pasquini2025llmmap}, and author attribution \cite{huang2025authorship}, employ principles of data traceability. Additionally, older best practices, i.e., logging\cite{Foalem2023}, also enter the domain of ML and AI.

Efforts to study traceability in LLMs are impeded by the inherently black-box nature of many models and the limited access to their architectures and parameters.
However, open-weight models enable insights into the model's parameters, creating opportunities to trace circuits of parameters \cite{circuit-tracer} through the LLM's generation process, and establishing openings for transparent and trustworthy inference.

\subsection{Transparency}

\begin{quote}
``Transparency is a multifaceted concept that is commonly used to refer to the act of \textit{being open}. Its use and interpretation depend on the context.'' \cite{Ofem2022}\\
``I define software transparency as a condition that all functions of software are disclosed to users.'' \cite{Meunier2008}
\end{quote}

We observe transparency in LLM systems as follows: 
Model parameters are disclosed and all additional features are observable by users, such as pre- \& post-filtering, auxiliary tool calls, weight and token distributions, as well as logits.
Furthermore, documentation options for LLM environments include dataset cards (dataset metadata), model \& system cards (model scope \& training characteristics).
Models themselves are clustered in two groups: (1) transparent, open-weight models with disclosed architecture and parameters, and (2) \textit{black-boxesque}, opaque, closed-weight models that restrict any access to their internals.


\subsection{Interpretability and Explainability}
\label{sec:interp-expl}

The terms \emph{interpretability} and \emph{explainability} are often conflated in AI literature.
Recent surveys suggest a growing trend to distinguish them more precisely across communities \cite{graziani2023global}, with complementary support from broader XAI reviews \cite{ali2023explainable}.
Graziani et al. \cite{graziani2023global} document that a distinction between interpretability and explainability is increasingly common among researchers and across disciplines, and Ali et al. \cite{ali2023explainable} provide additional evidence and a taxonomy that reinforces this emerging separation.

\begin{definition}[Interpretability]
\label{def:interpretability}
A \emph{passive} property of a model: the extent to which its internal structure, parameters, and mechanisms can be \emph{directly understood} by humans (e.g., small decision trees, sparse linear models, monotonic GAMs). Interpretability is typically achieved by design and allows reasoning about model behavior without auxiliary tools.
\end{definition}

When feasible, researchers generally prefer models or system components whose behavior can be understood by inspection. While LLMs with soaring parameter counts are not inherently interpretable, \emph{interpretable system design} can still be pursued: constrained modules, rule- or program-guided reasoning, monotonic submodels, or sparse decision layers. 
Interpretability is an inherent \emph{quality} of the model and often targets AI experts or auditors.

\begin{definition}[Explainability]
\label{def:explainability}
An \emph{active} process (often post hoc) that produces \emph{human-understandable} reasons for a model’s predictions or behavior, especially when the model itself is not transparent (e.g., feature attributions, counterfactuals, example- or concept-based explanations, procedural traces).
\end{definition}

For complex or opaque models, we present \emph{explanations} of particular outputs or global behavior. The focus is on \emph{human} understanding (including average users): why and how a result was produced; what steps or decisions were taken; and which factors influenced the outcome. This is an \emph{active} process, ideally via interfaces that present post-hoc explanations (e.g., attributions, counterfactuals, exemplars, chain-of-thought \emph{summaries}/rationales appropriate for the audience), with attention to fidelity, stability, and usefulness.

We deem this distinction important because, beyond clarifying intent, using consistent terminology across social and technical sciences reduces cross-disciplinary ambiguity, improves auditability, and aligns with governance and reporting practices.


\section{Data Provenance \& Attribution}
\label{chap_dp}

\begin{center}
\begin{table*}
  \caption{Data Provenance \& Attribution}
  \label{data_prov_table}
  \scalebox{0.75}{
  \begin{tabular}{llll}
    
    \toprule
    \textbf{Method} & \textbf{Evidence Direction} & \textbf{Artifact Entity} & \textbf{Article}\\
    \midrule
    \textbf{Licensing \& Attribution} \\
    \hline
     Dataset Licensing & Direct & Provenance metadata objects & \cite{Longpre2024}\\
     Self Attribution & Indirect & Claim attr. metadata, & \cite{Huang2024}\\
      & Bidirectional  & MultiAttr \& PolitiCite & \cite{patel2024}\\
     Source Attribution & Direct & Eval. attr. artifacts (AIS) & \cite{Rashkin2022}\\
     Code Attribution & Indirect & Function decompilation \& call graphs & \cite{Macdonald2024}\\
      & Direct & Lexical, structural \& semantic features & \cite{Li2023}\\
     Code Vulnerability & Direct & ModelCards & \cite{bandara2024}\\
      & Indirect & Patch data, commit information & \cite{Wang2024} \\
    
    \textbf{Provenance Extraction \& Retrieval} \\
    \hline
     Biomedical Retrieval & Direct & Query and passage pairs & \cite{Xu2024DP}\\
     Data Processing \& Analysis & Direct & Raw \& refined data, config. files & \cite{Sun2024}\\
     Metadata Extraction & Indirect & Model metadata (schemas) & \cite{Tsay2022}\\
     Information Extraction & Direct & InstructIE dataset & \cite{Gui2024}\\
    
    \textbf{Data Generation \& Annotation} \\
    \hline
     Data Annotation & Direct & Semantic Data Dictionary  & \cite{Rashid2020}\\
      & Bidirectional & Generated annotations & \cite{Choi2024} \\
     Balanced Data Training & Indirect & Pretraining model checkpoints & \cite{Sun2024DP}\\
     Data Transformation & Direct & Scripts, logs, commands, C2Metadata & \cite{Song2022}\\
     Data Generation & Direct & Query, documents, labels & \cite{Frej2020}\\
      & Bidirectional & Q\&A, Attribution spans & \cite{Golany2024}\\
     Data Fusion and Active Learning & Direct & Multimodal data source metadata & \cite{Blasch2022}\\ 
      & Direct & Annotated lexicons and rules &\cite{mishra2023} \\

    \bottomrule
  \end{tabular}}
\end{table*}
   
\end{center}

The quality of any AI-based system, including LLMs, heavily relies on the quality of its data sources (see, for example \cite{MOHAMMED2025102549}). 
From the initial training data throughout fine-tuning and additional refinement steps, balanced, high-quality data is quintessential for well-performing language models. 
Using poorly balanced training data may result in highly biased output, an increased chance of hallucinations, or other unwanted behavior. 
The training data, however, is often concealed. 
On the one hand, model creators do not fully disclose their training processes and data for fear of knowledge drain and business goals. 
On the other hand, current LLMs are not designed with a focus on linking source data to generated outputs. 
Language models could be designed with a dedicated architecture focusing on the linkage between the initial source data and the final model outputs.
Data provenance applied to the domain of language models could be useful to shed light on these complex and opaque AI systems.
We group the works corresponding to this chapter into three categories
and list their evidence direction and artifact entity in Table~\ref{data_prov_table}.
An essential aspect of responsible LLM development concerns the valid (1) \emph{licensing} of (training) data sources, and the accurate \emph{attribution} of these sources to generated outputs.
To identify, trace, and restore the origins of data sources for LLMs, (2) \emph{Provenance extraction} and {retrieval} techniques can be applied, while also supporting the analysis of model behaviour.
(3) \emph{Data generation}, a core competence of LLMs, enables the creation of synthetic data samples, and \emph{data annotation} is applied to enrich raw data with semantic information.
We classify the provenance evidence into two directions: \emph{Direct} evidence stemming from raw data sources to be ingested into LLM systems \cite{Longpre2024, Rashkin2022, Li2023, bandara2024, Xu2024DP, Sun2024, Gui2024, Rashid2020, Song2022, Frej2020, Blasch2022, mishra2023}, \emph{Indirect} evidence which arises with the help of LLMs \cite{Huang2024, Macdonald2024, Wang2024, Zambrano2024, Tsay2022, Sun2024DP}, and \emph{Bidirectional} evidence featuring both provenance stemming from data and LLMs \cite{patel2024, Golany2024}.
Furthermore, the \emph{artifact entities} describe the modalities of \emph{how} the provenance information is stored.

\subsection{Licensing and Attribution}

Foundational LLMs are trained on a plethora of data, which are grouped in datasets, over multiple iterations. 
At later stages of an LLM, additional datasets are involved when fine-tuning or during alignment. 
Extra datasets in the form of test sets can also be involved when evaluating the previously trained LLM, but compared to the training datasets, these test sets do not have a direct impact on the parameters of an LLM. 
Up until recently, these datasets were exclusively composed by humans and published with corresponding licenses. 
However, with the recent rise of generative AI, more and more synthetic datasets are created with the use of LLMs. 
These synthetically generated datasets pose the risk of forwarding bias inherited and unintentionally containing proprietary content from the initial training data of host LLMs. 
Data attribution concerning LLMs refers to the identification, tracing, and documentation of which particular data sources were used to train, fine-tune, or align a language model and how outputs trace back to these sources. 
Tracing back outputs means grounding answers that stem from models to source documents, which contain information and knowledge that was used in constructing the answer.
We found two approaches for attribution - \emph{Provenance-preserving} and \emph{Provenance-inferring} data attribution. 
Provenance-preserving data attribution refers to an exact match of source information and generated answer, such as commonly used in Retrieval Augmented Generation (RAG) systems.
In RAG-based systems, answers are created by a prompt, which is used to retrieve relevant information from, e.g., a vector store, based on similarity matching. 
After the retrieval, the context is forwarded to an LLM. 
This context acts as information grounding, which can naturally be attributed to generated answers, simply by the fact that it was forwarded as an informational context to the question. 
For classic LLMs used for generative AI, a more complex procedure is necessary. 
To link a generated text, or token, to every possible source from the training data, a throughout tracing mechanism, or a mapping for each parameter group (circuit) to their respective training samples is required. 

On the other hand, provenance-inferring data attribution refers to looking for justifying sources post-hoc, e.g., via a web search, to then ground information of an already generated answer on these sources. 
Although this forms an appealing approach, these methods do not genuinely trace sources to results but simulate them by providing justifying sources. 
Furthermore, provenance-inferring data attribution is limited by the availability of information for citing claims, which might not be possible for proprietary data sources, yet exists as knowledge in the language model.



\subsubsection{Dataset Licensing}

In their work, Longpre et al. \cite{Longpre2024} systematically audited more than 1.800 text-based datasets and call for more transparent ways of handling datasets for the usage of LLMs. 
They investigate dataset licensing and attribution, highlighting the grey zone of unspecified licenses on public datasets, and their missing categorization to commercial or non-commercial usage. 
Furthermore, Longpre et al. consider the datasets' origins and usage as provenance metadata objects directly from sources. 
They emphasize a focus on legal and ethical usage of data and present dataset distributions regarding contained topics and involved languages, which can be investigated through their created data provenance explorer DPExplorer\footnote{https://www.dataprovenance.org/data-provenance-explorer}.

\subsubsection{Self Attribution}

Huang et al. \cite{Huang2024} propose START, a self-taught attribution framework that enables LLMs to enhance their attribution capabilities to mitigate hallucinations autonomously. 
This is achieved without the help of human-labeled data or external supervision. 
They evaluated their START on three open-source datasets: ASQA, ELI5, and StrategyQA. 
For their data synthesis, they propose a five-step protocol that involves the initial answer generation without citations, claim decomposition and combination, as well as the document generation, which is sourced by the refined claims. 
Finally, they use the attribution labels for each document as references for the initial response as the last step. 
For self-improving, they employ a warm-up, rejection sampling, fine-tuning, and fine-grained preference optimization phase.

Patel et al. \cite{patel2024} address LLM self-attribution for long-form QA over multiple documents and sources. 
To overcome the scarcity of annotated multi-source attributable data, they enhance existing QA datasets with attributions (MultiAttr) and present a multi-source attribution dataset with multi-paragraph answers (PolitiCite). 
With their dataset, Patel et al. try to mitigate propagated LLM errors by using expert-written answers and human-retrieved evidence instead of relying on LLM-generated answers. 
To create MultiAttr, they propose two strategies using few-shot prompting: Expanding short to long answers and generating additional positive and negative sources for preexisting QA pairs. 

\subsubsection{Source Attribution}

Source attribution examines how model-generated outputs can be reliably traced back to identifying evidence and how outputs remain faithful to their underlying sources.
Frameworks such as the Attributable to Identified Sources (AIS) \cite{Rashkin2022} formalize this process by evaluating whether generated claims are fully supported by their cited documents.
Rashkin et al. \cite{Rashkin2022} introduce the AIS, an evaluation framework for assessing whether outputs of natural language generation (NLG) models are supported by source documents.
They describe how AIS can be applied to different NLG tasks, such as conversational QA, summarization, and table-to-text generation.
Additionally, they highlight that extractive systems, due to their design of quoting verbatim information from documents, achieve higher AIS scores and are less likely to output hallucinations.
Furthermore, Raskin et al. present a two-stage annotation pipeline for the evaluation of model outputs.

\subsubsection{Code Attribution and Vulnerability}

Source code generation is one of the core domains of AI-generated data occurring in the majority of LLM benchmarks.
In terms of attribution, this opens another question about the involvement of source code in the LLM data pipeline. 
How much and which code sources are used to train modern language models? 
And along with this question, new issues emerge regarding licensing, authorship attribution, code vulnerabilities, and malware attribution. 
Li and Hong et al. \cite{Li2023} propose a feature set to differentiate between human- and LLM-authored source code by observing style, technical level, and readability of code. 
MacDonald et al. \cite{Macdonald2024} introduce two sources of expressive static features for static malware authorship attribution. Using LLM embeddings to capture features at the function-level based on decompiled source code, and a Graph Neural Network (GNN) approach that takes into account both function-level analysis as well as global structural level analysis, which performs best. 

Bandara et al. \cite{bandara2024} and Wang et al. \cite{Wang2024} propose insights into vulnerability aspects of source code management. 
They propose DevSec-GPT to tackle software container vulnerabilities by leveraging blockchain and generative AI, as a custom-trained Llama2 language model. 
They establish effort towards provenance and attribution by their end-to-end pipeline that is handled via model cards, which are stored in the blockchain for immutability. 
Wang et al. propose an automated data collection framework consisting of three modules to 1) untangle vulnerabilities with LLMs and static analysis tools, 2) extract multi-granularity dependencies, and 3) filter outdated artifacts via a trace-based approach. 
Furthermore, Wang et al. introduce their repository-level vulnerability dataset ReposVul, created through patch data, dependency trees, and trace-based commit histories, providing valuable provenance information within its process.



\subsection{Provenance Extraction \& Retrieval}

Provenance extraction and retrieval focus on identifying, tracing, and restoring the origins of data used in LLMs.
This process enables researchers and practitioners to analyze in more detail how data influences model behaviour, while ensuring accountability and reproducibility.
By retrieving provenance information, these methods enhance transparency and allow deeper insights into the data flow across the models' life cycle. 

\subsubsection{Biomedical Retrieval}
Xu et al. introduce BMRetriever, a series of dense text retrievers in various sizes using LLMs as backbones \cite{Xu2024DP}. 
They use Pythia \cite{biderman2023pythiasuiteanalyzinglarge} for their 410M and 1B parameter, Gemma \cite{gemmateam2024gemmaopenmodelsbased} for their 2B, and BioMistral \cite{labrak-etal-2024-biomistral} for their largest 7B parameter retriever.
Two phases of pre-training are presented by Xu et al.: An unsupervised contrastive pre-training on biomedical query-passage pairs and instruction fine-tuning with diverse labeled data, which include synthetically generated examples from LLMs (GPT-3.5). 
Through verifying 200 random samples of the generated synthetic examples in a human evaluation with medical students, they found no traces of misinformation or hallucinations.
However, this highlights the risk of using synthetically generated data in an unsupervised learning setting.

\subsubsection{Data Processing \& Analysis}
Sun et al. \cite{Sun2024} present a data preprocessing framework for pretraining foundation models.
Because the datasets used for pretraining are often manually curated and cumbersome to create, their proposed framework is aimed at unifying data preprocessing.
A preprocessing module, with operators at different data granularity levels, and an analyzing module to probe refined data, make up the core of their framework.
Sun et al. automatically evaluate their framework with ChatGPT and additionally train two GPT-2 models with and without their data preprocessing pipeline, and report remarkable performance increases in comparison to the baseline.

\subsubsection{Metadata Extraction}
Zambrano \cite{Zambrano2024} proposes zero-shot prompting strategies for data extraction in a legal domain using Llama3-70B and Mixtral-8x7B, showcasing the strong capabilities of LLM-based information extraction. 
The experimental setup consists of extracting the first trial court outcome and the subsequent court ruling from 400 manually annotated cases of the French Courts of Appeal. 
Tsay et al. \cite{Tsay2022} introduce a library (AIMMX) to extract AI model metadata from software repositories into flexible metadata schemas alongside an evaluation dataset containing 7998 public models.

\subsubsection{Information Extraction}
Gui et al. \cite{Gui2024} present a human-annotated bilingual instruction-based information extraction dataset (InstructIE) and a framework (KG2Instruction) using Knowledge Graphs to generate similar datasets. 
The framework's most notable feature is using preexisting relation triples and generating missing triples using preexisting IE models.

\subsection{Data Generation \& Annotation}

Data generation is a core competence of LLMs, especially modern state-of-the-art Generative Pre-trained Transformers (GPTs), and their generated outputs range from single to more than 100.000 tokens.
In theory, the output of LLMs is not limited to a certain threshold, but the generation limits typically stem from fair practical reasons.
These massive generative capabilities enable the creation of entire synthetic datasets in a short amount of time, compared to human-curated data. 
However, this method of data generation introduces new risks into the domain of data science, which are linked directly to data provenance. 
Language models may propagate bias into the generated data, which originates from their own training data, forwarding existing problems.
Without human supervision, model hallucinations in generated data are difficult to detect and pose a threat to further unsupervised data usage.
Additionally, this raises the question of who the author of LLM-generated data is: The prompt-issuer, the model creators, or the mix of influential training data? 
To enable answering the latter, provenance information must be preserved throughout the generation process, e.g., via data annotations in generated text or datasets.

\subsubsection{Data Annotation}

A crucial step for constructing new datasets is data annotation \cite{Choi2024}. 
It involves labeling and enriching raw data with semantic information to make it interpretable for machine learning, analytical, and language models. 
High-quality annotations are quintessential to ensure validity, reproducibility, and fairness in further processing tasks.
Traditional data annotation relies on manual human labeling, which is not only time-consuming and costly but also prone to inconsistency.

Recent advances introduce automated and semi-automated annotation methods, including frameworks based on ontologies that formalize data semantics for interpretability \cite{Rashid2020, Song2022}, LLMs that act as autonomous or assisting annotators \cite{Choi2024}, and the creation of balanced, traceable multilingual text corpora that support transparent model training \cite{Sun2024}.

Rashid et al. \cite{Rashid2020} introduce the Semantic Data Dictionary (SDD), a specification aimed at machine-interpretable data annotations that adds semantics to traditional data dictionaries. 
SDD increases traceability and interoperability by linking tabular data to ontology concepts and provides data provenance by including \textit{wasDerivedFrom} and \textit{wasGeneratedBy} relationships within annotations.

Choi et al. \cite{Choi2024} propose to use GPTs as autonomous data annotators to generate multilingual, synthetic labels (\textit{silber annotations}) from limited available human (\textit{gold annotations}). 
From a provenance perspective, this method creates a traceable annotation pipeline that links human-gold labels with LLM-generated silver.

\subsubsection{Balanced Data Training}
Sun et al. \cite{Sun2024DP} present FuxiTranyu, a multilingual language model trained on balanced data. 
The training corpus consists of balanced multilingual data from 43 different natural languages and 16 programming languages following a provenance-aware pre-training approach.
Sun et al. employ heuristic \& learned quality filters, deduplication processes, and meticulously report the data collection process.

\subsubsection{Data Transformation}
Song et al. \cite{Song2022} aim to improve traceability and reproducibility for data transformations with their C2Metadata system that captures transformation and provenance information in Structured Data Transformation Language (SDTL) as part of metadata.

\subsubsection{Data Generation}

Frej et al. \cite{Frej2020} present MLWIKIR, a Python framework to automatically generate large-scale Information Retrieval (IR) datasets in multiple languages based on Wikipedia. 
Their future work points to using these generated datasets for Deep Learning (DL) training and observing model effects for different languages.
Golany et al. \cite{Golany2024} investigate LLM-based data generation in the setting of source-grounded information-seeking dialogs. Their approach follows a semi-automated approach, where an LLM is used to generate dialog queries and responses, which are then verified by humans for attribution spans. 
Additionally, they present MISeD, a dataset of information-seeking dialogs from the domain of meeting transcripts, and compare fine-tuning with this semi-automated dataset to manual data.

\subsubsection{Data Fusion \& Active Learning}

In active learning, an ML model proposes the next data points that should be labeled to learn most effectively. 
The model issues requests for training examples on which it is most uncertain to a third entity, most often a human, for labelling these data points. 
Blasch et al. \cite{Blasch2022} investigate active learning with joint data for model training as fusion-based AI, and fusion tools for multimodal data fusion and transfer learning.
PyTAIL is a Python framework for interactive and incremental model training enhanced with human feedback for continuously evolving text presented by Mishra et al. \cite{mishra2023}. It exceeds classic active learning approaches by also suggesting new rules and lexicons for human validation. With their work, Mishra et al. emphasize human-guided data labeling and feature curation.

\section{Traceability \& Accountability}
\label{chap_trace}

\begin{table*}
  \caption{Traceability \& Accountability}
  \label{tab:trace}
  \scalebox{0.80}{
  \begin{tabular}{llll}
    \toprule
    \textbf{Method} & \textbf{Open-Weight Models} & \textbf{Closed-Weight Models} & \textbf{Article}\\
    \midrule
     ML-based Logging Practices & - & - & \cite{Foalem2023}\\
     Data Transformation Recovery & Code LLama & GPT4 & \cite{Lou2024}\\
     Contextual Semantic Parsing & BART & - & \cite{moradshahi2023}\\
     Model Prediction Attribution & LLama2-7B, LLama3-8B, Mistral-7B, & - & \cite{su2024}\\
     & Gemma-7B, BART-large, DEBERTA-large & &  \\
     & Flan-T5-large & &  \\
     Periodic Watermarking & - & GPT-3 text-embedding-002 API & \cite{Ye2025}\\
    \bottomrule
  \end{tabular}}
\end{table*}


This chapter explores techniques that enhance traceability across both data and model dimensions of LLMs, with each paper forming a complementary pillar of traceability. 
Table~\ref{tab:trace} provides information about the topics and involvement of LLMs.
Traceability is omnipresent in the domain of computer science, and the underlying concept of stacked function calling possesses the capabilities of observing execution flows. 
This enables tracing bugs or unexpected behaviors at runtime to backtrack the initial system error. 
Furthermore, programmers document execution steps in their code with logging statements that contain information on current stages and error messages. 
In the world of ML, traceability is tackled similarly to general software engineering. 
Logging statements are used to track program flows and verbalize error and failure codes, which can be used to trace root causes. 
In addition, the domain of LLMs is also capable of adapting logging practices. 

\subsection{Logging}

Although similar in spirit, logging practices in ML differ from general software engineering. 
Foalem et al. \cite{Foalem2023} investigate the differences in these two domains and highlight the most important deviations in logging practices. 
They analyze 110 open-source ML-based applications from GitHub\footnote{https://github.com/} and compare them to logging practices of various software projects in different programming languages, as well as Android applications. 
Their findings reveal that ML-based logging is less pervasive than software projects written in JAVA, C\#, C, or C++, but more pervasive than Android applications. 
Additional findings reveal that \textit{WARNING} and \textit{INFO} logging statements occur the most in ML-related scenarios, and although ML-specific logging libraries exist, standard logging libraries are still prevalent in ML projects. 
Through a qualitative and quantitative analysis, Foalem et al. show that the model \textit{training} phase contains the largest amount of logging statements, whereas the \textit{deployment} phase contains the least. Furthermore, data, model, and configuration management are the most important information artifacts that are logged by ML developers.

Following the practices of ML-based approaches, logging can be used similarly in LLM-based AI projects. 
Operational traceability can be achieved by logging model states during inference to increase transparency and reproducibility, and datasets and epochs with intermittent results can be logged during the language model training process to create auditable provenance information.  
This ties directly into the domains of metadata capturing and inference tracking to achieve advancements in data provenance measures. 

\subsection{Data Transformation Recovery}

Data transformations occur when data entries in datasets change over time. 
Tables in a data repository often undergo iterative changes when data scientists or machine learning developers fine-tune their pipelines. 
These changes are seldom documented, leaving a final dataset without documentation or a change log. 
Lou et al. \cite{Lou2024} investigate the reconstruction of missing data transformation changes for tables in ML pipelines by employing LLMs for their DATALORE framework. 
In their DATALORE framework, Lou et al. use LLMs to infer missing data transformations between table-based datasets to increase traceability and reproducibility. 
This procedure is split into three parts:
First, the related table discovery finds connections from base tables to all the tables that contain similar and related information by applying equi-join, semantic join, and transformation-based join techniques. 
Second, the related tables are joined to a unified table, which is then passed to the LLM alongside the augmented table. 
To improve inference with the language model, this is performed on just a subset of columns to then generate a potential transformation. 
If the LLM-generated transformation can change the table entries to the augmented table, the transformation is considered valid. 
Third, the table selection refinement step examines all transformations to find a subset of transformations that are necessary to create the augmented table. 
This is achieved by employing a greedy algorithm to find a solution with the least amount of transformations from the base tables to the augmented tables. 

This method highlights how LLMs themselves can act as reasoning agents to reconstruct data transformation derivation paths, which form an essential part of provenance tracking. 
Furthermore, this incorporates traceability by ensuring that provenance information of training data is transparent and reproducible.

\subsection{Contextual Semantic Parsing}

Moradshahi et al. \cite{moradshahi2023} address the challenge of maintaining semantic and annotation consistency across multilingual dialogue datasets. 
In their work, they propose a Contextual Semantic Parsing (CSP) model which is based on BART, and evaluate it on the RiSAWOZ, CrossWOZ, CrossWOZ-EN, and MultiWOZZH datasets. 
The model encodes only the latest agent and user utterances in combination with dialogue states, to avoid compounding errors for translations, without losing accuracy. 
Additionally, they propose an automatic dataset translation technique using machine translation, instead of expensive human annotations. Using this approach, they present two versions of the RiSAWOZ dataset in English and German with an accuracy within 11\% of the original human-annotated versions, showcasing the capabilities of their translation technique without relying on human annotations. 

Regarding traceability, this work directly contributes to transparent transformation procedures by ensuring that the meaning, structure, and state annotations in multilingual datasets can be traced and aligned faithfully across translations. 

\subsection{Model Prediction Attribution}

In their work, Su et al. \cite{su2024} introduce a WrapperBox framework aiming at faithful attribution of model decisions to training data by exploiting the inherent transparent nature of three classic models kNN, DT and k-means. 
The overarching goal of the WrapperBox framework is to wrap LLM representations in classic neural models to trace model predictions back to specific training samples. 
They investigated seven open-weight language models: BART-Large, DEBERTA-large, FLAN-T5-large, LLama2-7B, LLama3-8B, Mistral-7B and Gemma-7B with four evaluation criteria: Accuracy, Precision, Recall, and F1-Score.

WrapperBoxes are an essential step toward faithfully traceable data attribution and algorithmic recourse by tracing model decisions back to training samples. Thus, improving efforts regarding model-data attribution.

\subsection{Periodic Watermarking}

Ye et al. \cite{Ye2025} introduce TimeMarker, a periodic watermarking framework that is designed to enhance temporal traceability and enable copyright protection for LLMs deployed in cloud environments that are used within the context of Embedding as a Service (EaaS). 
Their framework enables the identification of model extraction timings across particular time intervals (sub-periods) by embedding distinct watermarks in the model's output. 
This allows the detection of \textit{when} a model extraction or misuse occurs.
TimeMarker employs adaptive watermark strengths by using techniques like information entropy and frequency-domain transformation to adjust watermark intensity. They tested their approach across multiple datasets, such as Ag News, Enron, Mind, SST2, Trec, and verified detection of model extraction attacks over various sub-periods, and showcased their good performance in comparison to static watermarking approaches.

From the perspective of traceability, TimeMarker provides temporal accountability and traceability that allows verification of usage timelines, detection of unauthorized extraction attempts, and maintenance of ownership transparency.

\section{Transparency \& Opacity}
\label{chap_trans}

\begin{figure}
  \centering
  \includegraphics[width=0.8\linewidth]{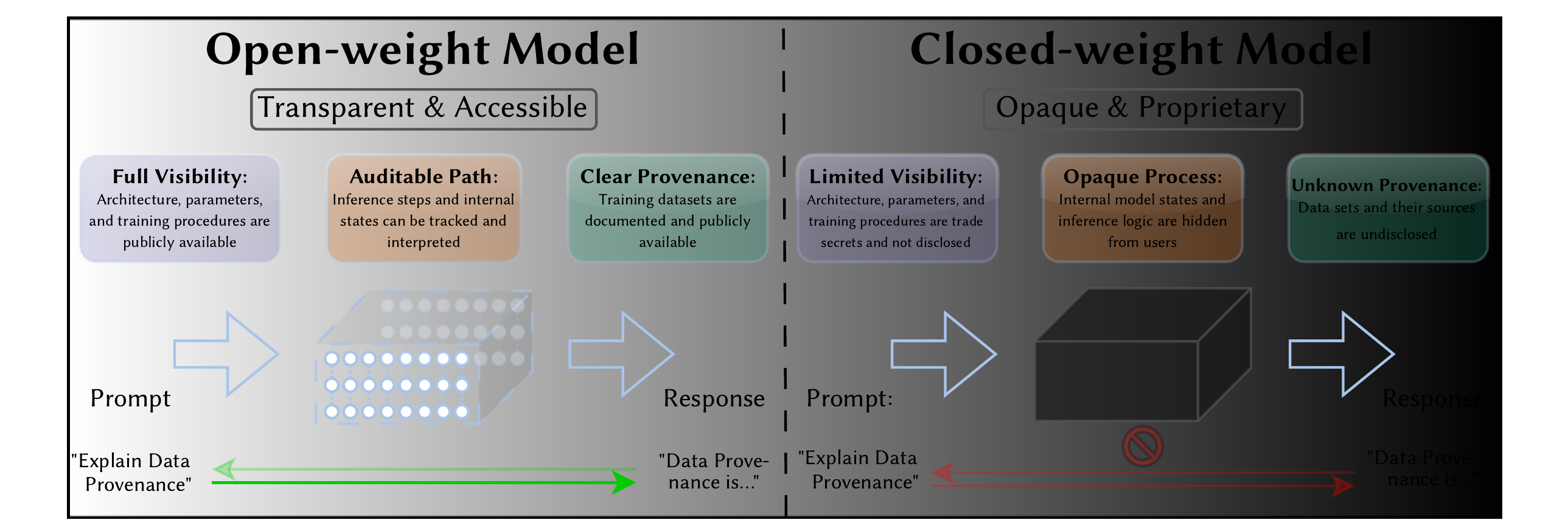}
  \caption{Differentiation between the key characteristics of open- and closed-weight LLMs, highlighting the difference in visibility, traceability, and data provenance.}
  \Description{Differentiation between the key characteristics of open- and closed-weight LLMs, highlighting the difference in visibility, traceability, and data provenance.}
  \label{open_closed_models}
\end{figure}

Transparency is emerging as a pillar in assessing the reliability and accountability of LLMs.
In regard to data provenance, transparency is a quintessential quality that enables the understanding of data flows and their origins.
A fundamental distinction divides current LLMs into two groups: Open-weight and closed-weight models.
Figure~\ref{open_closed_models} summarizes the architecture and key characteristics of open-weight and closed-weight language models.
Open-weight models are characterized by the disclosure of their architecture, parameters, and, in some cases, their training data, which allows for external scrutiny, transparent inference, and the possibility of investigating model inference on private hardware for fully traceable data flow. Table~\ref{tabel_openweight} summarizes the works incorporating open-weight models.
In stark contrast to these fully open models, closed-weight models can be described as black boxes, with related work outlined in Table~\ref{tabel_closedweight}.
Their intrinsic opacity restricts access to internal components entirely and provides little to no visibility into their data provenance, training procedures, and decision-making mechanisms. 
This hinders systematic evaluation and independent auditing while raising concerns about explainability, interpretability, and fairness. 
Current research on the transparency of LLMs examines how varying degrees of openness affect the interpretability and explainability of these systems. 
Consequently, the remainder of this chapter is split into four sections: (1) research using open-weight models, (2) works built upon closed-weight models, (3) human-centered interpretability opportunities of language models, and (4) a perspective on explainability capabilities.

\subsection{Open-weight models}

\begin{table*}
  \caption{Transparency work using open-weight models}
  \label{tabel_openweight}
    \scalebox{0.80}{
  \begin{tabular}{llll}
    \toprule
    \textbf{Topic \& Method} & \textbf{Open-Weight Models} & \textbf{Domain} & \textbf{Article}\\
    \midrule
     Short text similarity   & Pretrained sentence transformer models            & Computer Science       & \cite{Gagliardi2023}\\
     Prompt Learning         & BERT, T5, MT5, LLaMa-7B, Qwen-7B, Falcon-40B      & Computer Science       & \cite{Gu2024}\\
     Interpr. logic KBQA     & ChatGLM2, Baichuan2                               & Computer Science       & \cite{Xu2024}\\
     Social media analysis   & LLaMA-3, Sentence-BERT (+ GPT-3.5 turbo)          & Computer Science       & \cite{Ravi2024}\\
     KBVQ evaluation         & LLaVA-v1.5, InstructBLIP, MiniGPT-v2/-4, BLIP-2,  & Computer Science       & \cite{Cheng2025}\\
                             & VisualGLM, OpenFlamingo (+ GPT-4)                 &                        & \\
     Misspelling generation  & Own dense vector model (word2vec)                 & Biomedical Inf.        & \cite{Sarker2018}\\
     Computer-aided diagnosis& LLaVA, Llama 3.1                                  & Biomedical Inf.        & \cite{Mukhlis2024}\\
     Prompt engineering      & LLamA-3-70B, Mixtral 8x7B                         & Law \& Legal           & \cite{Zambrano2024}\\
     SAE-Model               & LLamA2-7B (+ GPT-4)                               & AI (NLP)               & \cite{Song2024Trans}\\
     LLM internal states     & LlamA2, Mistral-7B-inst.                          & AI (LLM)               & \cite{Ji2024}\\
     Meta classification     & InstructBLIP, mPLUG-Owl, MiniGPT-4, LLaVA-1.5     & AI (Multimodal)        & \cite{Fieback2025}\\
     CLIP image localization & CLIP                                              & Computer Vision        & \cite{Yao2024}\\
     Pixel understanding     & Vicuna LLMs, CLIP (+ GPT-4)                       & Computer Vision        & \cite{Yuan2024}\\
     Market Forecasting      & BERT, GPT-2                                       & Finance                & \cite{Zhang2024Laser}\\
     Meta learning           & BERT                                              & Cybersecurity          & \cite{Corona-Fraga2025}\\
    \bottomrule
  \end{tabular}
  }
\end{table*}

Open-weight models offer insights into their internal states and decision-making processes. 
This transparent design opens possibilities for responsible and interpretable LLM-based AI research.
Gagliardi et al. \cite{Gagliardi2023}, Gu et al. \cite{Gu2024}, Xu et al. \cite{Xu2024}, Ravi et al. \cite{Ravi2024}, and Cheng et al. \cite{Cheng2025} extend current research boundaries in the domain of Computer Science.

Gagliardi et al. \cite{Gagliardi2023} present a multilingual ensemble method that combines the benefits of open transformer embeddings and WordNet-based methods. 
The aim to evaluate the semantic similarity of short texts and keywords is explored across different approaches to identify similarities with different pre-trained language models.
In \cite{Gu2024}, the AGCVT-Prompt technique is introduced, which auto-generates Chain-of-Thought (CoT) rationales and verbalizers for prompt learning using models like BERT and T5. 
This approach significantly improves the transparency and interpretability of the decision-making process by employing dual verbalizers for the topic and sentiment prompt templates.
Furthermore, Xu et al. \cite{Xu2024} aim to improve the transparency of reasoning paths for Knowledge Base Question Answering (KBQA).
They combine fine-tuning LLMs (ChatGLM2 and Baichuan2) with the logic programming language Prolog for a generation-retrieval multi-hop KBQA method.
By enhancing the reasoning paths with the generation of logical forms (from Prolog), the reasoning becomes more interpretable.
In \cite{Ravi2024}, Ravi et al. present the PRAGyan framework, which integrates knowledge graphs (KGs) with LLMs to enhance the causal analysis in social media data (tweets).
Their approach employs retrieval-augmented generation (RAG) to retrieve contextually relevant information from a KG. 
Thereby, they utilize LLMs such as LLaMA-3 to infer causal relations in tweets by utilizing a KG, yielding more transparent and interpretable outputs.
Cheng et al. \cite{Cheng2025} present FacGPT, a scalable and deterministic GPT-2-based evaluator model for automatic KVQA evaluation that achieves better correlation with GPT-4 and human evaluators, increasing efficiency by 300\%. 
Additionally, they introduce the K-VQA dataset that aims to improve the evaluation of answer correctness by including human annotations.

The biomedical domain is often confronted with ambiguous text, stemming from illegible handwriting or misspellings.
If undetected, these mistakes are automatically forwarded into medical datasets, leading to multiple instances of the same medical term (e.g., medications). 
Sarker et al. \cite{Sarker2018} present a misspelling generation approach for health domain keywords using cosine similarity over vector representations and lexical similarity, which defaults as an unsupervised approach but can be enhanced with supervision. 
In their work, they employ their own dense vector model (word2vec) to enhance retrieval quality on often misspelled medical terms.
Missing interpretability hinders the transparency of LLM-based computer-aided diagnosis (CAD).
In \cite{Mukhlis2024}, Mukhlis et al. present a novel CAD framework that enhances CAD, with LLM-powered visual and textual interpretability.
Their method enhances the transparency between the medical staff, the CAD system, and end-users by creating a trustworthy environment. 
In their work, they propose an eight-step CAD procedure that utilizes visual and textual interpretability from the LLaVA and Llama 3.1 LLMs they employ.

Zambrano \cite{Zambrano2024} evaluates two open-weight LLMs: Llama-3-70B and Mixtral-8x7B for legal outcome extraction in a zero-shot prompt setting.
The study uses a dataset of 400 manually annotated legal decisions from the French Courts of Appeal.
Furthermore, Zambrano shows that accessible open-weight models can achieve close to state-of-the-art accuracy without fine-tuning.
Its results highlight that open-weight LLMs paired with prompt engineering can serve as transparent, high-performing alternatives to closed-weight proprietary LLMs for data extraction in the legal domain.

Furthermore, AI research itself is investigated in regards to SAE-models \cite{Song2024Trans}, LLM internal states \cite{Ji2024}, and multimodal systems \cite{Fieback2025}. 
Song et al. \cite{Song2024Trans} present a Slot-Aware Enhanced Model (SAE) to improve zero-shot dialogue state tracking (DST) in small pre-trained models (LLaMA2-7B), tackling the closed-source nature and inaccessibility for local deployment of advanced LLMs. 
Furthermore, generating slot-aware questions using GPT-4, and using this synthetic data to train a small-scale model.
The trustworthiness and reliability of LLMs are significantly limited by their risk of hallucinations.
In \cite{Ji2024}, Ji et al. investigate the internal states of LLMs in the context of hallucinations. 
Similar to humans, who possess the ability to recognize missing knowledge when presented with a question, internal mechanisms of LLMs can provide the same information. 
Their analysis reveals two key insights: (1) internal states reveal whether an LLM has seen/not seen a query during training, and (2) internal states show the likelihood to hallucinate when presented with a query. 
Ji et al. investigate neurons, activation layers, and tokens and achieve an accuracy of 84,32\% on hallucination detection at runtime. They employ LLaMA2-7B and Mistral-7B as their LLM subjects for internal state detection. 
Fieback et al. \cite{Fieback2025} present MetaToken, a binary classifier for hallucination detection on the token-level in image descriptions with meta classification. 
The classifier can be applied to any open-source LVLM without requiring ground truth information. The presented meta-classification approach decides whether a prediction is true or false based on uncertainty features of LLMs. 
Additionally, they evaluate their proposed method on four state-of-the-art LVLMs: InstructBLIP, mPLUG-Owl, MiniGPT-4 and LLaVA-1.5, outperforming confidence-based and prompting baselines.

Computer Vision has become a critical component of modern AI systems, enabling models to process, interpret and generate visual information \cite{Yao2024, Yuan2024}. 
Yao et al. \cite{Yao2024} present an image localization method for single images based on CLIP-guided prompts.   
The weights of the integrated score map enhance the model's transparency and improve interpretability. 
CLIP and clear guidance mechanisms provided by prompts clarify feature identification and extraction.
Thus, improving the general interpretability and transparency as well as reliability.
Yuan et al. \cite{Yuan2024} present Osprey, a mask-text instruction tuning approach that extends MLLMs with a CLIP backbone. 
To achieve this, regions are masked into language instructions aiming at pixel-wise visual understanding. 
Yuan et al. employ GPT-4 to generate high-quality mask-text pairs for their Osprey-724k dataset. 
Furthermore, reporting State-of-the-art region understanding and reduced hallucinations.

Also, the financial domain is supplemented by transparent AI research, aiming toward interpretable market forecasting \cite{Zhang2024Laser}, 
In \cite{Zhang2024Laser}, Zhang et al. introduce a BERT-based long and short-term memory retrieval (LASER) architecture to forecast market volatility as well as tackle interpretability concerns with their GPT-2-guided BEAM algorithm. 
The BEAM architecture generates human-readable narratives that verbalize the evidence and path leading to the model prediction. 
Together, these two approaches form the LASER-BEAM pipeline.

Finally, toward cybersecurity, Corona-Frage et al. \cite{Corona-Fraga2025} propose a QA framework for vulnerable source code review that integrates BERT with a Prototype-Based Model-Agnostic Meta-Learning (Proto-MAML) approach. 
BERT is leveraged with Few-Shot Learning (FSL) to detect and explain code vulnerabilities with minimal data, yet achieving F1 and EM scores of close to 100\%. 
By combining BERT's embeddings with meta-learning adaptability, their approach produces transparent and interpretable outputs, which enhances traceability in AI-assisted code analysis.

\subsection{Closed-weight models}

\begin{table*}
  \caption{Transparency work using closed-weight models}
  \label{tabel_closedweight}
    \scalebox{1.0}{
  \begin{tabular}{lllc}
    \toprule
    \textbf{Topic \& Method} & \textbf{Closed-Weight Models} & \textbf{Domain} & \textbf{Article}\\
    \midrule
     Transparency of LLMs        & ChatGPT, Bard/Gemini      & Phil. \& Epist. of AI  & \cite{Heersmink2024}\\
     Educational AI              & GPT-4                     & Education                         & \cite{Misiejuk2024}\\
     GenAI for grading           & GPT-4o, GPT-4-turbo       & Education                         & \cite{Thomas2025}\\
     Online log analysis         & Vicuna-13B, GPT-3.5-turbo & Computer Science                  & \cite{Liu2024}\\
     LLM-based decomposition     & GPT-3 Codex               & Computer Science                  & \cite{Ye2023}\\
     LLM on medical reports      & GPT-3/-3.5/-4             & Medicine                          & \cite{Wei2024}\\
    \bottomrule
  \end{tabular}
  }
\end{table*}


Closed-weight language models are perceived as black-boxes.
This goes along with unwarranted attitudes of trust and trustworthiness \cite{Heersmink2024}. 
Heersmink et al. \cite{Heersmink2024} investigate this topic from a philosophical and epistomological standpoint. 
They propose that this stems from a lack of data transparency and algorithmic transparency by analyzing the current phenomenon of anthropomorphizing of LLM systems in the form of chatbots such as ChatGPT and Bard.

AI and LLMs are setting foot in teaching, Misiejuk et al. \cite{Misiejuk2024} and Thomas et al. \cite{Thomas2025} explore their impact on the domain of education. 
They \cite{Misiejuk2024} examine the potential of LLMs like GPT-4 for coding student discourse, highlighting the importance of transparency for generative AI in the educational domain and presenting moderate reliability and barriers such as high cost and token limits. 
Similar to the topic of educational AI, Thomas et al. \cite{Thomas2025} present the effectiveness of multiple-choice Questions relative to open-response questions. 
They compare the performance of 234 tutors and investigate two LLMs, GPT-4o and GPT-4o-turbo,  for automatic grading of open-response questions, highlighting that prompting for rationale can help in obtaining more interpretable outputs.

The domain of computer science itself is also prominent in LLM research - Liu et al. \cite{Liu2023} and \cite{Ye2023} investigate log analysis and table-based reasoning.
Liu et al. \cite{Liu2023} present LogPrompt, an interpretable log analysis approach for online scenarios using LLMs (GPT-3.5 \& Vicuna) and a suite of enhanced prompting strategies, showing large performance gains (380\%) in comparison to using single prompts. 
A human evaluation of the interpretability of LogPrompt yields high scores for the usefulness and readability of the generated content and justifications. Additionally, LogPrompt also supports the employment of smaller-scale open-source LLMs.
Ye et al. \cite{Ye2023} present an approach to employ LLMs (GPT-3 Codex) for decomposing evidence and questions for table-based reasoning. 
Additionally, they highlight that their method even surpasses human performance on the TabFact dataset and provides interpretability of results via tracing of generated sub-evidence and sub-questions.

Finally, Wei et al. \cite{Wei2024} present an evaluation framework for LLM performance on medical questions, based on reports. 
Within their meta-analysis, ChatGPT achieved an accuracy result of 56\% in addressing medical queries.

\subsection{Interpretability}
\label{sec:interpretability}

\begin{table*}[t]
  \caption{Human-centered interpretability and model passive characteristics}
  \label{tabel_closedweight}
    \scalebox{0.90}{
  \begin{tabular}{lllc}
    \toprule
    \textbf{Characteristic \& Method}  & \textbf{Article} \\
    \midrule
     Explanation rendering approaches & \cite{ko2024} \cite{Gu2024} \cite{ZhaoLFTQA} \cite{Lu2020} \cite{Yang2022} \cite{Kim2024} \cite{shenZei}\\
     Concept- and representation-level methods and domain-specific considerations & \cite{Patrcio2025} \cite{Williams2024} \cite{Zhang2023} \cite{Yuan2025} \cite{YuLi2024} \cite{Ma2025}\\
     Architecture, hybrids, and editing & \cite{Bangerter2024} \cite{Zhang2024KA} \cite{Hocking2022} \cite{Yang2025}\\
     Causal and evaluation-driven interpretability & \cite{Rodriguez-Cardenas2023} \cite{Burns2020} \cite{Ruga2024}\\
     A philosophical aside: do only the activated weights “matter”? & \cite{elhage2022superposition} \cite{chan2022causal} \\
    \bottomrule
  \end{tabular}
  }
\end{table*}

Interpretability provides insight into how model inputs relate to outputs, making it a key component of transparency. 
We use \emph{interpretability} in a user-centric sense: the degree to which intended users can understand \emph{why} a model produced a given output in their task context, with minimal cognitive load. 
In modern deep/LLM systems, this is typically operationalized by (i) designing modules that promote understandable intermediate structure, and (ii) post-hoc procedures that render behavior legible for different audiences.

\subsubsection{Explanation rendering approaches.}
One line of work seeks to surface intermediate reasoning so users can follow the path to an answer. \emph{Trace-/Chain-of-Thought} style prompting decomposes complex problems into subproblems, exposing stepwise rationale to enhance interpretability of the process rather than only the final answer \cite{ko2024, Gu2024}. 
Plan-first pipelines similarly externalize plans before execution to make long-form reasoning auditable (e.g., planner–reasoner–verbalizer for LFTQA) \cite{ZhaoLFTQA}. 
While such rationales aid human understanding, they require fidelity checks to ensure they track the model’s actual computation.

Token-/feature-level views remain central. 
Layerwise Relevance Propagation variants adapted to sequential student modeling \cite{Lu2020} and specialized biomedical attributions/visualizations (e.g., BertViz-based analyses of circRNA models) \cite{Yang2022} illustrate how local importance signals and circuit overviews help users form mechanistic intuitions. 
For LVLMs, tools like Logic Lens visualize layerwise behaviors to detect when and where reasoning emerges, addressing transparency and reproducibility concerns \cite{Kim2024}. 
More broadly, interpretability benefits from \emph{UI scaffolding}: explicitly structured reports, checklists, and reviewer prompts that make the \emph{process} of generating and validating summaries transparent to humans \cite{shenZei}.

\subsubsection{Concept- and representation-level methods and domain-specific considerations.}
A complementary line constrains or probes models via \emph{human-interpretable concepts}. 
Concept Bottleneck Models (CBMs) promise inherent interpretability, with predictions being mediated by named concepts, but face annotation burden and concept availability; recent work mitigates this with pretrained VLMs and LLMs to propose or validate concepts \cite{Patrcio2025}. 
Topic-modeling work shows that richer, human-facing outputs (beyond token lists) improve perceived reliability and interpretability for readers \cite{Williams2024}. 
For domain knowledge, Markov Logic Networks paired with LLM-assisted knowledge acquisition/formalization target transparent, checkable intermediate structure \cite{Zhang2023}.

Across application areas, ``interpretability" is pursued via domain-tailored structure plus human-facing outputs: txt2onto 2.0 leverages LLM embeddings to make unstructured biomedical metadata legible via ontologies \cite{Yuan2025}; interpretability is treated as a first-class metric alongside performance in entity–relation extraction under noise-robust training \cite{YuLi2024}; and in HR analytics, fine-tuned LLMs are evaluated not only on accuracy but also on the clarity of their rationales relative to classic ML baselines \cite{Ma2025}.

\subsubsection{Architecture, hybrids, and editing.}
Hybrid systems balance raw flexibility and clarity.
Examples include ANFIS+LLM hybrids for social-media analysis \cite{Bangerter2024}, interpretable transfer frameworks (e.g., KAI / ZSSD) for zero-shot learning \cite{Zhang2024KA}, and white-box surrogates or sparsified layers to expose decision rules in safety-critical settings \cite{Hocking2022}. 
Targeted model editing offers an interventionist perspective via modifying hidden states or components to align behavior with human expectations and then auditing the localized effect. 
This approach is explicitly framed as being interpretability-oriented \cite{Yang2025}.

\subsubsection{Causal and evaluation-driven interpretability.}
Causal perspectives aim to separate signal from spurious correlation. 
Galeras introduces a causal-inference benchmarking frame across software-engineering tasks to interrogate LLM behavior \cite{Rodriguez-Cardenas2023}. 
Statistical tests like the Interpretability Randomization Test (IRT) and One-Shot Feature Test (OSFT) probe whether purportedly important features actually matter, by systematically substituting or randomizing inputs/features and observing degradation \cite{Burns2020}. 
In regulated domains, preference for simpler statistical models is often motivated by \emph{auditability and justification} rather than accuracy alone; recent surveys argue for prioritizing interpretable modeling (and privacy-aware pipelines) where stakes are high, while exploring LLMs for logical reasoning as complementary aids \cite{Ruga2024}.

\subsubsection{A philosophical aside: do only the activated weights “matter”?}
A tempting assumption is a kind of \emph{compartmentalization}: for a given prediction, the currently activated units/weights are the ones that ``really" matter, suggesting an inherent, passive interpretability of deep nets. 
Mechanistic work on \emph{superposition} and \emph{polysemanticity} shows why this is unreliable: networks often pack many unrelated features into overlapping directions, so the same unit can implement multiple, context-dependent features; causal contribution is not cleanly localized to what happens to be firing strongly \cite{elhage2022superposition}. 
Methods like \emph{causal scrubbing} test mechanistic hypotheses by resampling or swapping activations along putative circuits and checking whether behavior is preserved, offering a more principled way to validate “which parts mattered” beyond raw activation magnitude \cite{chan2022causal}.


\subsection{Explainability}


\begin{table*}
  \caption{Explainabilty}
  \label{tab:explainability}
  \scalebox{0.85}{
  \begin{tabular}{llllc}
    \toprule
    \textbf{Overal Goal} & \textbf{Scope} & \textbf{Stage} & \textbf{Application} & \textbf{Article}\\
    \midrule
      &  &  Post-hoc & Model Agnostic & \cite{Arabzadeh2024} \cite{Bhaskar2024} \cite{Lu2023}\\

     Generating explanations using LLMs & Local  &  Post-hoc & Model Specific & \cite{Yu2024} \cite{Shah2025}\\
       &  &  Ante-hoc & Model Specific & \cite{YangMental}\\
    \midrule

     Augmenting existing explanations & Local and Global &  Post-hoc & Model Agnostic & \cite{Valina2024}\\
        \midrule

     Explaining LLMs & Global &  Post-hoc & Model Agnostic & \cite{Topal2025}\\

    \bottomrule
  \end{tabular}}
\end{table*}

Explainability focuses on generating reasons for a model's decision-making process, thereby enhancing transparency. 
Following the definition in Section~\ref{def:explainability}, explainability is an active process focused on producing human-understandable rationales for a model's decision-making process. 
In this process, developed artifacts have different characteristics based on their scope, inference stage, goal, and relation to the underlying model. 
Literature on eXplainable AI (xAI) \cite{graziani2023global} \cite{ali2023explainable} typically defines local and global approaches as the scope of explanations provided by the explainability technique. 
Local explanations are focused on explaining the behavior of a specific instance or input to the model, whereas global explanations provide rationales for the whole model's behavior and remain true to overall observations. 
Additionally, literature outlines post-hoc and ante-hoc xAI method categories as characteristic of a stage of model development in which these methods are implemented. 
Post-hoc methods refer to the methods employed after the model has been trained, while ante-hoc methods are deployed during the training and development stages. 
Finally, model-specific and model-agnostic are characteristics of a method related to its connection to the underlying model it is trying to explain. 
While model-specific methods are tied to mechanisms of specific models, model-agnostic methods (often also post-hoc) are independent of the kind of model used. 

Several overarching goals were identified and analyzed in the following section through an in-depth analysis of the identified literature related to explainability. 
Outlined in Table \ref{tab:explainability}, alongside the previously described characters, these overarching goals are (1) generating explanations using LLMs, (2) augmenting existing explanations, and (3) explaining LLMs. 

\subsubsection{Generating explanations using LLMs}
Literature with the first overarching goal, \emph{generating explanations using LLMs}, utilizes LLMs to generate local explanations for a specific downstream task. 
Arabzadeh et al. \cite{Arabzadeh2024} develop a framework for evaluating the applicability and utility of LLM-powered applications. 
In their work, they construct a framework consisting of several agents, such as \emph{Critic Agent, Quantifier, and Verifier}, which, based on the nature of the task, develop criteria and evaluate the utility of the applications.
They suggest that analyzing the behavior of such agents contributes to the explainability of the overall utility evaluation task. 
Following a similar concept, Bhaskar and Stodden \cite{Bhaskar2024} propose \emph{ReproScore} metrics evaluated through an LLM. 
Although the overall task is predictive, the analysis of the evaluation steps helps understand the overall decision-making process. 
Finally, in \cite{Lu2023}, Lu et al. aim to improve the explainability through a few-shot biomechanical knowledge fusion framework. 
The presented approach incorporates hierarchical context into prompts, leveraging models' reasoning capabilities while respecting hierarchical constraints to make knowledge base integration more transparent. 

While previously outlined methods are model-agnostic, Yu et al. \cite{Yu2024} present a model-specific process for LLM fine-tuning to provide explainable recommendations for Click-Through Rate prediction. They developed a custom reward function - an alignment process to reflect user intentions, and utilize the LLaMa-3-7b model for finetuning using a \emph{Low-Rank Adaptation} approach. 
Further research \cite{Shah2025} explores using a GPT-4-based language-based explainability component to generate explanations of model behavior in a cancer detection downstream task. The component utilizes activation feature maps and GradCAM visualization to generate high-level textual explanations. 
Finally, contrary to previously outlined approaches, where methods are utilized in a post-hoc manner, Yang et al. \cite{YangMental} present ante-hoc, model-specific methods for generating explanations in evaluating mental health on social media.
Through the fine-tuning process, these models are presented with explanations and evaluations from mental health experts and adapted to generate explanations jointly with predictions.  

\subsubsection{Augmenting existing explanations}
Work focused on the second overarching goal, \emph{augmenting existing explanations}, aims to enhance, transform, or adapt existing explanations to make explanations more accessible and interpretable. 
Recent research \cite{Valina2024} introduced a method based on established xAI techniques LIME and SHAP augmented with natural language capabilities of LLMs, to address the challenges of augmenting existing explanations. 
The proposed approach uses these xAI methods to analyze the contribution of different features in deep generative models. 
Subsequently, this post-hoc method proposes generating user-adapted, natural language explanations suited to end users with varying levels of technical expertise. 

\subsubsection{Explaining LLMs}
Finally, \emph{explaining LLMs}, focuses on understanding the characteristics of the model itself. 
Following this idea, Topal et al. \cite{Topal2025} developed a set of LLM evaluations with an emphasis on trustworthiness, fairness, and robustness. 
These evaluations provide an insight into the global characteristics and capabilities of several Turkish-capable LLMs. 
Overall, the multi-prompt-based method was evaluated on models such as: Llama-3.1-8B-Instruct, Qwen2.5-7B-Instruct, Mistral-7B-Instruct, gemma2-9b-it, making the overall approach model-agnostic.

\section{Bias \& Uncertainty}
\label{chap_bias}


\begin{table*}
  \caption{Bias \& Uncertainty}
  \label{tab:commands}
  \begin{tabular}{llc}
    \toprule
    \textbf{Topic} & \textbf{Focus} & \textbf{Article}\\
    \midrule
     Meassurig bias & Gender bias in text data& \cite{Jain2021}\\
     Data curation  & Amplifying bias & \cite{Taori2022}\\
     LLMs in research  & Amplifying data and model bias & \cite{Jain2023}\\
     LLMs for text classification & Mitigating human-induced bias & \cite{Fuhnwi2025}\\
     Quantifying uncertainty & Input alteration & \cite{Hou2024}\\
     Data and model uncertainty & Software engineering & \cite{Sallou2024}\\
     Performance and Reliability & Scaling laws & \cite{Cherti2023}\\
    \bottomrule
  \end{tabular}
\end{table*}


Bias and uncertainty analysis reveal how the underlying training data and model processes influence the system's behavior, highlighting the need for provenance information to detect and mitigate harmful outputs.
\textbf{Bias} is inherent in human behavior \cite{emberton2021unconscious} and, consequently, it is in data. 
This might lead to detrimental bias or discrimination in algorithmic results, decreasing ``fairness". 
As such, research on understanding sources of bias, quantifying bias, and methods for mitigating bias has been conducted in the general context of AI-based systems (e.g., \cite{ntoutsi2020bias}), as well as LLMs in particular (e.g., \cite{gallegos2024bias}).

When bias is present in LLM training data, the bias is adopted by the model itself. Bias and discrimination can then non-insignificantly guide the model's actions.
\citet{Jain2021} propose an approach that exploits methods from NLP and Explainable AI to calculate a gender bias score based on the common co-occurrence of gendered nouns and pronouns in close textual proximity, i.e., within sentences. 
When unregulated, the models learn directly from data, and the stronger the bias in that data, the stronger it will be in the model's output. 
Consequently, a reinforcement effect is expected when LLM training data is scraped from the Internet, because more and more text is generated by LLMs themselves. 
When model outputs replace human annotations, feedback loops occur that lead to generalization issues and the probable amplification of bias, e.g., through less diverse content or newly introduced biases. 
The investigation of emerging data feedback cycles suggests that it is essential to keep LLMs well calibrated, prefer the application of sampling-based decoding strategies, and keep humans in the loop of data generation \cite{Taori2022}.
Similar feedback loops are to be expected in scientific writing.  
While the application of LLMs opens the publication sphere to non-native speakers at a rapid pace, commonly used LLMs include all kinds of bias, particularly in failing to depict inclusion and diversity efforts, which is nowadays considered good practice in scientific communities. 
This can be partly attributed to the training on historical data sources. 
Additional bias is introduced by user preferences, e.g., through up or down voting, or by providing user context or specific prompting that the algorithm adapts to. 
Thus, if LLMs are trained on AI-generated research content, biases and errors will increase, and unwanted or outdated patterns are likely to propagate and repeat \cite{Jain2023}.
However, other research endeavors also explore the linguistic capacity of LLMs to mitigate human bias in text and classification tasks, ultimately improving automation. 
\citet{Fuhnwi2025} for instance, investigates the use of LLM-generated categorization of spam data to avoid bias from humanly hand-labeled spam messages and proposes an approach that succeeds with a fairness and accuracy increase.

\textbf{Uncertainty} is a quantifiable construct that indicates the degree to which an LLM is confident in the accuracy, robustness, and bias of its output. 
The level of uncertainty plays a pivotal role in user trust. If uncertainty is not marked properly, people may be irritated by the LLM's hallucinations or, even worse, rely on incorrect information.

Commonly used as a reliability indicator, it can be calculated in two components: 
(1) \textbf{Aleatoric (data) uncertainty}: which results from ambiguity in data or prompting and more (unknown unknowns), and (2) \textbf{Epistemic (model) uncertainty}: which increases if the model lacks skill or information to answer a question \cite{wang2025aleatoric}.

\citet{Hou2024} introduces a framework for uncertainty decomposition to evaluate LLMs on a more informed level. 
By exploiting alterations to input prompts, a strategy that is applicable for black-box models, the research work elicits different manifestations of uncertainty, enabling quantification of individual components and the construct as a whole. 
\citet{Ji2024} investigates whether LLMs' internal states can be exploited to determine uncertainty by inferring the risk of hallucination in response to a provided query. 
The study concludes that LLMs' self-assessment, which takes into account internal states, shows more promise than other state-of-the-art methods.


\citet{Fieback2025} present an approach towards predicting hallucinations in large visual language models. Independent of the ground-truth dataset, the method is based purely on model output analysis, performing meta-classification to identify mistakes, i.e., hallucinated objects.

Other works address the underlying factors that increase a model's uncertainty. 
For instance, uncertainty may increase due to data leakage between training data and memorization. 
This likely leads to generalization issues and biased performance evaluations, which become apparent when the model is confronted with novel questions. 
Furthermore, there is no guarantee that outputs remain consistent over time, as temperature changes or model updates may cause deviations, which, again, lead to uncertainty due to the failure of reproducibility results \cite{Sallou2024}.
One manifestation of a lack of reproducibility appears when semantically similar queries yield inconsistent or inaccurate answers (non-deterministic). 
\citet{Yang2025} investigates this phenomenon and proposes a method that exploits the artificial addition of noise or bias to the hidden layer of the most important components of the LLM. 
The authors claim to have successfully improved consistency in semantically deviating queries.  

A more general approach to increasing model reliability and performance is to train according to scaling laws that aim to define the optimal ratio between model complexity, dataset size, and computing power. 
Recently, \cite{Cherti2023} investigated the accuracy of scaling laws using open-source data and models. 
They conclude that, in addition to dataset size, the specific characteristics of data are particularly influential on model and task-specific scaling. 
Thus, for the success of further research and development, foundation datasets and open datasets will play a pivotal role.

\section{Privacy}
\label{chap_privacy}

Data Privacy is a concern when using LLMs - especially if there is no access to the underlying model~\cite{Sallou2024}.
The relationship between privacy and provenance warrants a deeper analysis.
Differential privacy~\cite{dwork2006differential} is an approach that addresses membership inference attacks.
In machine learning, the objective is to prevent individual data points from being identifiable within the training set. 
In other words, the trained model should behave similarly whether or not a specific instance is included in the data. 
At the surface level, this goal appears to contrast with that of data provenance.

Some approaches aim to combine privacy preservation with key elements of data provenance.
In \cite{Kirmayr2025}, Kirmayr et al. propose a system for an in-car assistant, where users may have concerns regarding the use of private data.
They highlight that one should distinguish between semantic memory and episodic memory in conversational systems.
To accomplish a desired level of privacy, users may opt out of specific classes that are utilized in an information extraction step.
They further propose that transparency can be achieved by showing what data has been stored and where.

The use of access control to ensure protection of private and confidential data is presented by Jiang et al. \cite{Jiang2024}, in a clinical context.
Furthermore, they highlight the need for interpretability of systems in such a setting.
To this end, they examine hierarchical RAG systems in which multiple indices are governed by distinct access control rules. 
Searches are distributed across the indices for which the current user has read permissions, and the retrieved results are subsequently aggregated. 
Their work demonstrates that RAG-based architectures can serve as a practical means of achieving transparency and data provenance.

\section{Provenance Tools \& Techniques}
\label{chap_tools}

Provenance tools and techniques are the methodological backbone for capturing, managing, and analyzing the data flow throughout the LLM life cycle. 
Starting at the origin of data sources operationalized during training to the generated output of resulting models, tools \& techniques enable sense-making of LLM decisions and tracing truthful attribution of answers to underlying data references, depending on the transparency potential of the chosen models.
We separate tools from techniques by the following characteristics: (1) Tools are reusable libraries, datasets, visual inspection suites, or end-to-end systems, and (2) Techniques are modeling or methodological approaches that can be implemented across multiple tools and domains.
Furthermore, we present important tools \& techniques found throughout our literature review and summarize key concepts for each of the extracted components.
Some works could be listed as either a tool or a technique. For overlapping items, they are placed based on their apparent main contribution.

\subsection{Tools}
\label{subsec:tools}

\citet{Roberts2022}, \citet{Yadagiri2024}, \citet{Shah-Mohammadi2024}, \citet{iranfar2024}, and \citet{Longpre2024} present provenance tools that aid in capturing and analyzing data flows through LLMs.

\begin{itemize}
  \item \textbf{{t5x} and {seqio}.}
  Open source libraries for JAX and TPU-based training and for task-based data pipelines. The design supports deterministic processing, reproducibility, recoverability, and provenance tracking \cite{Roberts2022}. 

  \item \textbf{Direct Velocity for authorship verification.}
  A BERT-based classifier with linguistic features for human versus AI attribution. Useful for integrity checks and compliance verification \cite{Yadagiri2024}. 

  \item \textbf{Automated meta analysis pipeline.}
  Outcome extraction with GPT-4 and semantic outcome alignment with Sentence BERT. The pipeline standardizes heterogeneous clinical reports and improves traceability \cite{Shah-Mohammadi2024}. 

  \item \textbf{Surgical skill assessment dataset with soft data.}
  The Vertex Pursuit task pairs expert free text comments with interaction logs. Transparency comes from documenting the data pipeline and methods, while traceability is ensured through structured metadata with timestamps, labels, and comments. \cite{iranfar2024}. 

  \item \textbf{Data provenance explorer}
  The data provenance explorer enables users to trace and filter data provenance information of popular finetuning data collections with an interactive UI. The explorer improves dataset transparency and responsible data use \cite{Longpre2024}.
\end{itemize}

\subsection{Techniques}
\label{subsec:techniques}

This section outlines the techniques presented by \citet{shankar2019}, \citet{Zhu2024}, \citet{TongYu2024}, \citet{Macdonald2024}, \citet{Feuer2024}, and \citet{circuit-tracer} that can be applied across multiple tools and domains.

\begin{itemize}
  \item \textbf{Legal query reformulation as monolingual NMT.}
  Encoder \& decoder with attention to rewrite noisy legal queries into well-formed queries using semi-supervised pairs from logs. The paper documents the data generation pipeline, preprocessing steps, and reports on the algorithm used and the results obtained, covering several important parts of the data provenance.  \cite{shankar2019}. 

  \item \textbf{Statement level code summarization with LLMs.}
  Evaluation of GPT-4, GPT-3.5, CodeLlama, and StarChat against CodeT5 on Java and Python, including prompt and temperature analyses. The paper explains the provenance of the dataset, covering selection, processing, evaluation, statistics, model configurations, experimental setups, and both metric-based and human-assessed results.  \cite{Zhu2024}. 

  \item \textbf{Knowledge graph augmented diagnosis with LLM prompts.}
  The DeLL approach fuses EMR text with knowledge graph disease subgraphs encoded with GCNs and uses LLM prompts to enrich representations for prediction. By documenting dataset provenance, detailing the preprocessing pipeline, including de-identification, and presenting the system architecture, the study enhances both traceability and transparency. Traceability is achieved through clear data lineage and transformation steps, while transparency is supported by openly describing design choices and privacy safeguards, enabling reproducibility and ethical evaluation. \cite{TongYu2024}. 

  \item \textbf{Trace-of-Thought distillation and task decomposition.}
  Transfer teacher reasoning to smaller students through structured traces and decompose questions into steps to expose intermediate reasoning. This work enhances reasoning transparency by breaking problems into structured steps, enabling human-in-the-loop oversight and correction before execution. \cite{Macdonald2024}. 

  \item \textbf{Prior Data Fitted Network scaling with TuneTables.}
  TuneTables is a parameter-efficient fine-tuning method that scales Prior-Data Fitted Networks (PFNs) like TabPFN to millions of samples using soft prompt tuning. It promotes fairness via demographic parity regularization and improves interpretability by summarizing key features, enhancing transparency and trust.\cite{Feuer2024}. 

  \item \textbf{Circuit tracing}
  Enables a deeper understanding of model decision-making by observing circuits of parameters during model inference.
  Includes a tool to trace circuits\footnote{https://github.com/safety-research/circuit-tracer} \cite{circuit-tracer}.

\end{itemize}


\section{Discussion}


This section synthesizes our key findings across existing literature on data provenance, transparency, and traceability for LLMs, highlighting emerging trends, methodological challenges, and open questions that have shaped recent years, current progress, and the future landscape of AI development. 
In addition to that, it comprises the growth of adjacent domains of interpretability, explainability, bias \& uncertainty, provenance tools \& techniques, as well as data privacy, which are closely tied to the discipline of data provenance for LLMs.

As the language model parameter size keeps rising, new approaches for data acquisition are needed, and preexisting pitfalls in data handling become more pronounced.
From the error-prone licensing of datasets \cite{Longpre2024} to the generation of missing annotations in databases \cite{Choi2024}, the problem of scarce data labeling and unstructured, widespread data sources arises.
Data \cite{su2024} and source \cite{Huang2024, patel2024} attribution emerge as the clear goal of trustworthy, reliable AI development.

Best practices in software engineering, such as logging \cite{Foalem2023}, also emerge in domains such as LLMs, generative AI, and ML. Furthermore, new approaches such as watermarking \cite{Ye2025} and self-attribution \cite{Huang2024} appear.

Current model developments split AI research into two directions: (1) trust-focused, transparent, and open interpretation of language models, and (2) feature-driven, opaque, performance development. 
Then, we distinguish interpretability as a user-facing understanding of why a model produced an output (through inherent, passive characteristics of the model), and explainability as the process and interface that generates human-readable reasons. 
This separation helps align evaluation and governance. 
It also clarifies when to prefer interpretable design choices and when to rely on post hoc methods. 
We note that explainability-focused work aims for human-centered understandability, while interpretability-driven research deals with complex model analysis.

Inherently, human bias is imparted into data sources that form the foundation of training data for LLMs. This leads to language models adopting unexpected behaviours \cite{Sallou2024} and biased outputs \cite{Jain2021}.
Provenance tools \cite{Longpre2024} and techniques \cite{circuit-tracer} emerge towards AI data governance. 
Data privacy, however, remains an open question and a pressing concern. 
Due to the one-directional, final flow of knowledge into models, it appears impossible to protect one's personal information, especially with little to no insight into the training data. 
But, approaches like model editing \cite{Yang2025} aim to surgically remove undesired information from language models.

To summarize, data provenance goes hand in hand with transparency and traceability for LLMs. 
Without a traceable data flow, the origins of individual datums cannot be retained, and without a transparent, open model design, traceability is seemingly impossible. 
Bias is intrinsic to data sources and thus propagates into language models' training data. 
Data privacy concerns are directly linked to provenance tools and techniques, which aim to tackle the rise of data obfuscation in connection to LLMs.

\section{Current Limitations \& Future Work}

\begin{itemize}

    \item \textbf{Observation of classes in training data}: In some cases, training data involves classes that represent types of knowledge. Future work should investigate whether these classes also propagate within model parameters, and their origins should be traceable and attributable. 
    \item \textbf{Data contamination}: LLMs are trained on public information corpora gathered from the web, and the sheer amount of data makes manual filtering impossible. It is unknown whether evaluation datasets are also included within the training corpora. Future research should investigate what impact of this spillage and the mixture of the test data on the evaluation of LLMs.
    \item \textbf{Propagating biases through language models}: Biases are intrinsic to human-authored data sources. These biases propagate into LLMs through training and can lead to unintended and unwanted behaviour. Using LLMs for data generation forwards the \textit{learned}, biased characteristics into synthetic data sources. Future research should aim to detect and mitigate biases resulting in LLM responses for artificial data generation. 
    \item \textbf{Data provenance for LLM-generated content}: LLMs' generative capabilities have become popular for the creation of data. This source of synthetic data raises questions about the provenance of information, such as: Who is the author of such datasets?
    Future research should aim at a deeper understanding of the implications of these data transformations for LLM-based data generation.
    \item \textbf{Scarcity of labeled, annotated data}: Manually labeling and annotating datasets is expensive and time-consuming. We found that some reviewed works propose using LLMs to carry out the labeling step, also in a semi-supervised method. Future research should investigate what impact these LLM-generated, synthetic data labels have on the scientific community.
    \item \textbf{Prompt and version drift in LLM pipelines}: Reproducibility relies on stable prompts and LLM settings and versions. Small changes to prompts or silent model updates or changes can change outputs and derived labels. This weakens traceability and provenance claims - leading to possible unstable explanations.
    \item \textbf{Faithfulness of explanations}: Natural language explanations and attribution maps may not reflect the true reasoning behind model predictions. Future work should include causal validation of explanations through controlled interventions and perturbation analyses to assess fidelity and faithfulness.
    \item \textbf{The right to be forgotten}: The right to be forgotten describes the right to remove one's personal information from online directories accessible through internet searches. LLMs, as well, can unknowingly ingest private data and personal information into their parameters that one wishes to erase. Future research should investigate methods to \textit{detect} and \textit{delete} private information from language models, e.g., model unlearning.
 
\end{itemize}
\section{Conclusion}

This survey offers a thorough overview of the developments in data provenance, together with transparency and traceability for LLMs over the past decade.
Along the supporting pillars, bias \& uncertainty, privacy, and provenance tools \& techniques, we summarize key findings and highlight interconnections.
Data provenance for LLMs demands attention not just for data flows into LLMs, but simultaneously also for output data generated by LLMs. 
Transparency enables the investigation of models' decision-making and understanding procedures, whereas a lack of transparency may hinder it.
Traceability aims at ways to follow the trails of data and its knowledge propagation through LLMs, with logging and watermarking emerging as noteworthy approaches.
The surveyed papers make clear that without an active process, provenance information is being lost through the parameterization of model training in LLMs and the generated output data can no longer be attributed to the original influencing sources.
Bias is inevitably present in data and also leads to bias in LLMs.
This bias can be mitigated to a certain extent, but according to literature cannot be fully avoided.
Privacy concerns highlight the need to be able to trace personal information being embedded into LLMs, with a lack of effective methods highlighting a relevant research gap.
The survey presents provenance tools and techniques, designed to aid in tracing provenance information.
They help in unraveling complex model decision-making processes to understand LLM-internal procedures.
%
To synthesize the considered works, tracking the origins of data sources flowing into LLMs will enable the attribution of model decisions and generated responses to underlying sources, and will increase transparency and reproducibility.
Finally, this survey lays out a roadmap of further research directions and future work toward transparent, robust, and reproducible AI research for LLMs.

\section{Acknowledgements}
This work has been supported by the FFG, Contract No. 915297: “TrustInLLM” and Contract No. 911655: "Pro²Future II is funded within the Austrian COMET Programme (Competence Centers for Excellent Technologies) under the auspices of the Austrian Federal Ministry of Innovation, Mobility and Infrastructure (BMIMI), the Austrian Federal Ministry of Economy, Energy and Tourism (BMWET), and of the Provinces of Upper Austria and Styria. COMET is managed by the Austrian Research Promotion Agency FFG.”

\bibliographystyle{ACM-Reference-Format}
\bibliography{__main__}
\end{document}